\documentclass[preprint,aps]{revtex4}

\usepackage{amsmath,amssymb,amsfonts,dcolumn,color,graphicx,graphics,setspace,latexsym,setspace,lscape,subfigure,placeins,epsfig,hyperref}

\newtheorem{lem}{Lemma}[section]
\newtheorem{pro}[lem]{Proposition}

\newcommand{\be}{\begin{equation}}
\newcommand{\ee}{\end{equation}}
\newcommand{\ba}{\begin{eqnarray}}
\newcommand{\ea}{\end{eqnarray}}

\newcommand{\4}{\frac}

\newcommand{\lb}{\left}
\newcommand{\rb}{\right}

\def\bs{\begin{split}}
\def\ess{\end{split}}
\def\be{\begin{equation}}
\def\ee{\end{equation}}
\def\bea{\begin{eqnarray*}}
\def\eea{\end{eqnarray*}}
\def\f{\frac}

\begin{document}

\title{\Large \bf On higher order geometric and renormalisation group flows}

\author{Kartik Prabhu, Sanjit Das and Sayan Kar}
\email{kartikprabhu.iitkgp@gmail.com, sanjit@phy.iitkgp.ernet.in, 
sayan@iitkgp.ac.in}
\affiliation{\rm Department of Physics and Meteorology {\it and} Centre for Theoretical Studies \\Indian Institute of Technology, Kharagpur, 721302, India}

\begin{abstract}
\noindent Renormalisation group (RG) flows of the bosonic nonlinear $\sigma$-model
are governed, perturbatively, at different orders of $\alpha'$, by
the perturbatively evaluated $\beta$--functions. In regions where
$\frac{\alpha'}{R_c^2} << 1$ ($\frac{1}{R_c^2}$ represents the curvature
scale) the flow equations at various orders in $\alpha'$ can be
thought of as {\em approximating} the full, non-perturbative RG flow.
On the other hand, taking a different viewpoint, we may consider the
abovementioned RG flow equations as viable {\em geometric} flows in their
own right and without any reference to the RG aspect. 
Looked at as purely geometric flows where higher order terms
appear, we no longer have the perturbative restrictions (small curvatures). 
In this paper, we perform our analysis from both these perspectives
using specific target manifolds such as $S^2$, $H^2$, unwarped $S^2\times H^2$
and simple warped products.

\noindent We analyze and solve the higher order RG flow equations
within the appropriate perturbative domains and find the {\em corrections} 
arising due to the inclusion of higher order terms. Such corrections,
within the perturbative regime, are shown to be small and 
they provide an estimate of the error which arises when higher orders
are ignored. 

\noindent 
We also investigate the higher order geometric flows on the same manifolds
and  figure out generic features of geometric evolution, 
the appearance of singularities and solitons. The aim, in this context,
is to demonstrate the role of the higher order terms in modifying the
flow. One interesting aspect of our analysis 
is that, separable solutions of the higher order flow equations
for simple warped spacetimes (of the kind used in 
the bulk-brane models with a single extra dimension), 
correspond to constant curvature Anti-de Sitter (AdS) 
spacetime, modulo an overall
flow--parameter dependent scale factor. The functional form of
this scale factor (which we obtain) 
changes on the inclusion of successive higher order terms in the flow.  
\end{abstract}

\pacs{04.20.-q, 04.20.Jb}

\maketitle
\section{Introduction and overview} 

In the 1980s, Friedan \cite{friedan} first analysed the renormalisation group (henceforth referred as RG) flows 
for nonlinear $\sigma$-models and showed how one could arrive at the
vacuum Einstein equations $R_{ij}=0$ if one assumed quantum conformal (Weyl)
invariance, thereby equating the metric $\beta$-functions to zero. This became
a crucial result in the context of string theory and a route towards
connecting/obtaining Einstein's general relativity with/from string theory. 
Several authors, subsequently proved important generalisations \cite{sen} 
(particularly in the context of heterotic string theory),
which provided further evidence of the abovementioned connection. 
The fact that general relativity (modulo some additions) 
does emerge from string theory from the requirement of quantum
conformal invariance, is a more or less accepted fact, today.

Around the same time in the 1980s, Hamilton \cite{hamilton}, 
in the context of mathematics,
proposed the Ricci flow equations, with a largely different motivation,
primarily related to the geometrisation conjecture of Thurston, which
involves the topological classification of three manifolds.
More recently, Perelman's work \cite{perelman} on Ricci flows and its utility
in proving the Poincare conjecture has drawn many researchers 
to revisit such geometric flows\cite{Chow}. 

It is well--known that the Ricci flow 
equations are the same (modulo a proper scaling of the 
flow parameter) as the first order RG flow equations of
the nonlinear $\sigma$--model. However, one needs to
keep in mind the fact that the RG flow equations are obtained
from the perturbative evaluation of the $\beta$--function and thus, they
{\em approximate} the full, non--perturbative (and as yet unknown)
flow in the small curvature regime (i.e. $\frac{\alpha'}{R_c^2}<<1$,
where $\frac{1}{R_c^2}$ is related to the curvature scale) \cite{polchinski}. 
On the other hand, from a different perspective, the same equations can be thought
of as {\em geometric flows} where higher order terms are 
consistently included. We shall perform our analysis  from both these
points of view, in this article.

Recently, various aspects 
(eg. irreversibility, gradient flow) of $\sigma$-model RG flows 
(with and without the dilaton and Kalb-Ramond field), at the lowest order
have been investigated
from a mathematical/geometrical perspective \cite{oliy}. 
The connection between RG and geometric flows have been nicely
discussed in \cite{carfora}, \cite{bakas}. In particular, Carfora \cite{carfora}makes a valid point about the connection through the following statement
in his article: {\em `the structure of Ricci flow singularities suggests a 
natural way for extending,
beyond the weak coupling regime, the embedding of the Ricci flow into
the renormalization group flow'}.
 
Going beyond the lowest order, the authors in \cite{oli,oli1} and \cite{streets} have analysed 
second order RG flows in some detail by looking at
gradient flow aspects, monotonicity etc. 
Earlier, Tseytlin \cite{tseytlin},
claimed to have proved the monotonicity of the $\sigma$-model RG
flow to all orders in $\alpha'$ though some important objections
(at the second order level) regarding monotonicity were pointed out in
\cite{oli1}.

In our work here, we revisit 2nd order flows for
various simple target manifolds and also analyse
further higher order flows.
We first treat the flow equations as geometric flows 
in their own right,
and figure out the consequences thereof by looking for
general solutions, solitons, singularities etc. Subsequently,
we point out the behaviour of the flows in the 
perturbative, RG flow domain
and investigate the effect of including the higher order terms. 

It is important to mention the classes of manifolds we consider
in our discussion. These are essentially of two kinds--unwarped ones
and warped product varieties. In the former case, we consider the
simplest toy examples -- namely the two sphere and
two dimensional hyperbolic space. We look at the higher order
flow equations and find exact solutions, from which we can arrive at useful
conclusions. We also investigate an unwarped product briefly.
Later, we look at warped product manifolds where the generic line element is
assumed to be that of a bulk spacetime in a braneworld model with
one extra dimension \cite{brane}. For all the manifolds under consideration
gradient flow and monotonicity, upto second order in $\alpha'$, holds
without any problems \cite{oli1}.  

Thus, our main focus in this work is two-fold:

$\bullet$ Obtaining the perturbative corrections that arise due to the
inclusion of higher order terms in the RG flow equations.

$\bullet$ Studying the general geometric evolution, solitons, singularities
etc. which arise when we treat the RG flow equations at various
orders, as genuine
geometric flows in their own right, without reference to the perturbative
RG aspect. 

\section{ The RG flow equations at various orders}

The RG flow equations for the background 
metric coupling $g_{ij}$,  in bosonic non-linear $sigma$-models are given by --
	\be\label{RG flow}
		\frac{\partial g_{ij}}{\partial \lambda} = - \beta_{ij}
	\ee
where, $\beta_{ij}$ are the \emph{beta-functions}, usually obtained 
order by order, perturbatively. In a field theory
context, the parameter $\lambda$ is related to the momentum cut-off
scale $\Lambda$. Cut-off independence of the theory at length scales
larger than inverse $\Lambda$ gives rise to the above RG flow
equation.   

As mentioned earlier, the $\beta_{ij}$ is usually obtained at various orders 
--i.e. as a series in the parameter  $\alpha '$ (proportional to the inverse of the
string tension). More specifically, we may write,
	\be\label{beta expansion}
		\beta_{ij} = \alpha ' \beta^{(1)}_{ij} + {\alpha '}^2 \beta^{(2)}_{ij} + {\alpha '}^3 \beta^{(3)}_{ij} + {\alpha '}^4 \beta^{(4)}_{ij} + O\lb( {\alpha '}^5 \rb)
	\ee
where the various terms in the R. H. S. of the above equation, for
the metric nonlinear $\sigma$-model, are given by:
	\be\label{beta 1}
		\beta^{(1)}_{ij} = R_{ij}
	\ee
	\be\label{beta 2}
		\beta^{(2)}_{ij} = \frac{1}{2}R_{iklm}{R_j}^{klm}
	\ee
	\be\label{beta 3}
		\begin{split}
			\beta^{(3)}_{ij} = 	& \frac{1}{8}\nabla_p R_{iklm}\nabla^p{R_j}^{klm} - \frac{1}{16}\nabla_i R_{klmp}\nabla_j R^{klmp} \\
							   	& + \frac{1}{2}R_{klmp}{R_i}^{mlr}{{R_j}^{kp}}_r - \frac{3}{8}R_{iklj}R^{kspr}{R^l}_{spr}  
								\end{split}
	\ee
	\be\label{beta 4}
		\begin{split}
			\beta^{(4)}_{ij} = 	& -\4{1}{16} R_{1} +\4{1}{48} R_{2}-\4{1}{16}[\4{1}{2}+\zeta(3)] R_{3}+\4{1}{4}[1+\zeta(3)] R_{4}\\
							   	& +\4{1}{16}[\4{13}{3}-3\zeta(3)] R_{5}+\4{1}{8}[\4{2}{3}-\zeta(3)] R_{6} +\4{1}{4}[\4{8}{3}+\zeta(3)] R_{7}+\4{1}{4}[-\4{1}{3}+\zeta(3)] R_{8}\\
								& +\4{1}{12} R_{9}+\4{1}{12} R_{10}-\4{1}{6} R_{11}+\4{1}{16}[\4{4}{3}+\zeta(3)] R_{12}-\4{1}{4}[\4{4}{3}+\zeta(3)] R_{13}+~\mathrm{higher ~derivatives}		\end{split}
	\ee\\	
with $R_{1}$ to $R_{13}$  representing various combinations of 
Riemann tensors(for explicit expressions see \cite{jack}). The above expressions are valid, in the RG flow context,
only if $\frac{\alpha'}{R_c^2} << 1$ (where $\frac{1}{R_c^2}$ 
is the curvature scale).

We rescale the flow parameter $\lambda \rightarrow \lambda \alpha '/2$ to bring the leading order term of Eq.(\ref{RG flow}) in the form of a Ricci flow. Thus we now have the modified flow equations as --
%	\be\label{RG flow mod}
%		\frac{\partial g_{ij}}{\partial \lambda} = - 2R_{ij} - \alpha ' R_{iklm}{R_j}^{klm} +  \ldots
%	\ee
\be\label{RG flow mod}
		\frac{\partial g}{\partial \lambda} = - 2Rc - \alpha'  {\check{Rc}}^{(2)}-2\alpha'^2{\check{Rc}}^{(3)}- 2\alpha'^3 {\check{Rc}}^{(4)} \ldots
	\ee
where ${\check{Rc}}^{(2)}$ is a symmetric $2$ tensor defined as: ${\check{Rc}}^{(2)} = R_{iklm}R_{jabc}g^{ka}g^{lb}g^{mc} $
, ${\check{Rc}}^{(3)}=\f{1}{2}R_{klmp}{R_i}^{{}{mlr}}{{R_j}^{kp}}_{r}-\f{3}{8}R_{iklj}R^{kspr}{R^l}_{spr}+...+..$
 and ${\check{Rc}}^{(4)} =\beta^{(4)}_{ij}$
or, in component form, we have,
\be\label{RG flow mod1}
		\frac{\partial g_{ij}}{\partial \lambda} = - 2R_{ij} - \alpha'  R_{iklm}{R_j}^{klm} -  \ldots
	\ee
The abovementioned higher order terms are fairly difficult to obtain and 
involve laborious calculations. We have quoted the above results from 
articles in
\cite{sen,jack}. As far as we know, the terms beyond 4th order for
the metric $\beta$-function are not available in the literature.			
In the following, we will analyse the above flow equations
both from the RG and geometric flow perspectives. We also note that
knowing the geometric flow does give us useful knowledge about
the domain of validity of the RG flow, which is the
domain of low curvature.  

It may be asked, why we retain the coefficients in the RG flow equations 
while discussing higher order geometric flows. It is true that these
coefficients could be arbitrary when the RG flow aspect is ignored. 
However, the gradient and monotonicity 
aspects of the flow
is largely governed by the explicit values of these coefficients. 
Hence, even though flows may be defined with arbitrary coefficients
we stick to the RG flow values in our analysis of the higher
order geometric flows. 

We mention that we have done our analysis for both
$\alpha'>0$ and $\alpha'<0$. In the $\sigma$-model context, $\alpha'<0$
has no meaning (recall that $\alpha'$ is related to the inverse
string tension). However, one may consider $\alpha'<0$, while treating
the equations as those for geometric flows.  
	
%%%%%%%%%%%%%%%%%%%%%%%%%%%%%%%%%%%%%%%%%%%%%%%%%%%%	

\section{Higher order flows on surfaces}
In this section, we explicitly solve the flow equations for 2nd, 3rd and
4th order flows, using $S^2$, $H^2$ and $S^2\times H^2$ as toy
examples. The intention here is to figure out the role of the higher
order terms in changing the nature of the geometric flow. 
Consider  $(M, g)$ where $M$ is the two dimensional surface(spherical or hyperbolic)  and $g_{M}(\lambda)$ is the metric on $M$ in the  form:\\
\be\label{}
g_{M}(\lambda)= x^{2}(\lambda)~g^{can}_{M}
\ee
Here $x^{2}(\lambda)$ is the scale factor and $g^{can}_{M}$ is either the 
canonical spherical or the hyperbolic metric, depending on the case studied. 
We again rescale $x^{2}$, $\lambda $ as $\alpha' x'^{2}$ and $\lambda'\alpha'$ respectively for $\alpha'>0$. We also follow the same rescaling for $\alpha'<0$ except that,
we change $ \alpha'$ to $-|\alpha'|$. In order to distinguish, 
we will use the scale factor $a(\lambda)$ and $b(\lambda)$ for spherical and hyperbolic surfaces, respectively. After rescaling, the general equation for $x'^{2}(\lambda')$ is found to be:\\
\be\label{gensolsurf1}
 \f{dx'^2}{d\lambda'}=-2k-\f{2}{x'^2}~-k~\f{5}{2}\f{1}{x'^4}~-~2(\4{59}{24}\zeta(3)+\4{29}{24})\4{1}{x'^6} , \textrm{for $\alpha'>0$} 
 \ee
 \be\label{gensolsurf2}
  \f{dx'^2}{d\lambda'}=-2k+\f{2}{x'^2}~-k~\f{5}{2}\f{1}{x'^4}~+~2(\4{59}{24}\zeta(3)+\4{29}{24})\4{1}{x'^6} , \textrm{for $\alpha'<0$}
\ee
 Here $k=\pm 1$ stands for spherical (+) or hyperbolic (-) surfaces. 
The first term in the R.H.S corresponds to unnormalized Ricci flow whereas 
the subsequent terms are for 2nd, 3rd and 4th order flows, respectively. 
 \subsection{Geometric flow analysis}
\subsubsection{\underline{2nd order flow on $S^2$}}
Recall that the Ricci flow on the two--sphere gives rise to an ancient
solution. Including the second order term leads to the following proposition. 
\begin{pro}\label{innpd}
For $2nd$ order flow on the canonical two-sphere, with any choice of initial radius  and $\alpha'<0$ we obtain $a'_{\infty}=1$. If $\alpha'<0$ and $a'_0=1$ we obtain a soliton, while for $\alpha'>0$ and with any $a'_0$, we
obtain an ancient solution. 
\end{pro}

\begin{figure}[htbp] 
\centering
\subfigure[$a'_0= 1, \lambda'_{s}=0.153$,$\alpha'>0$]{\includegraphics[width=0.32\textwidth]{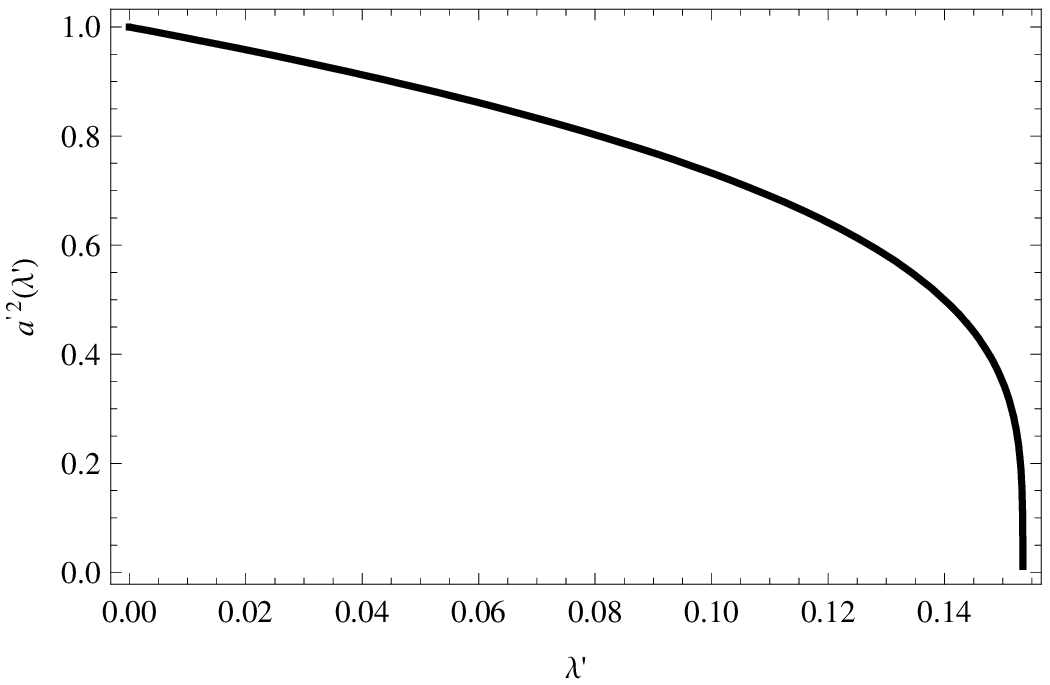}}
\subfigure[$a'_0= 1$, $\alpha'<0$, soliton]{\includegraphics[width=0.32\textwidth]{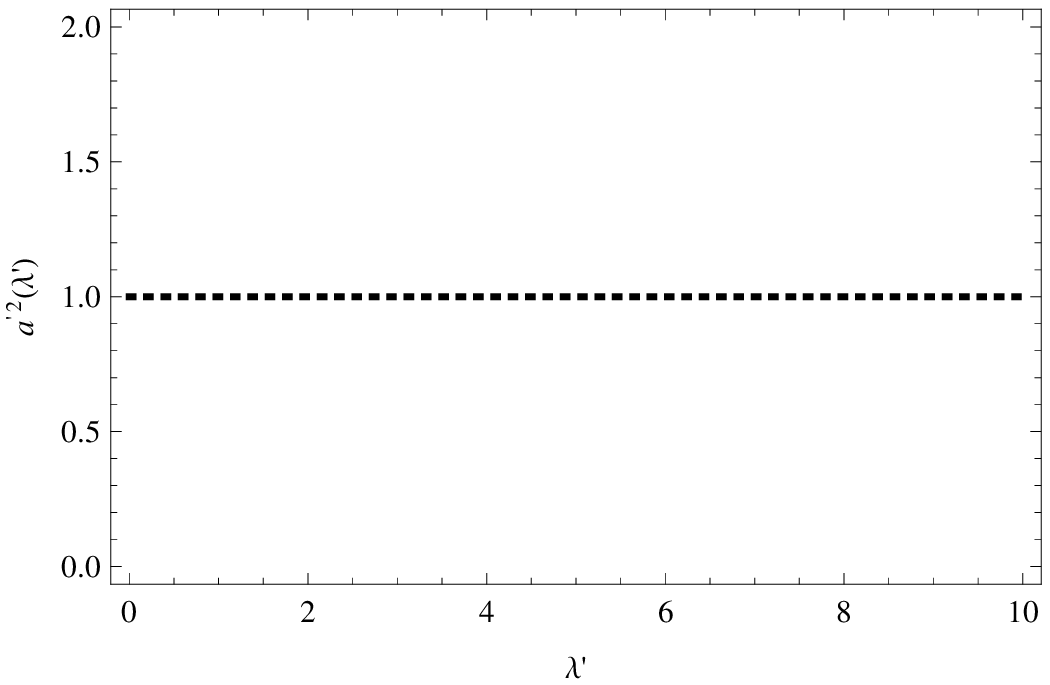} }\\
\subfigure[$a'(\lambda')^2>1,a'_0= 6$, $\alpha'<0$,]{\includegraphics[width=0.32\textwidth]{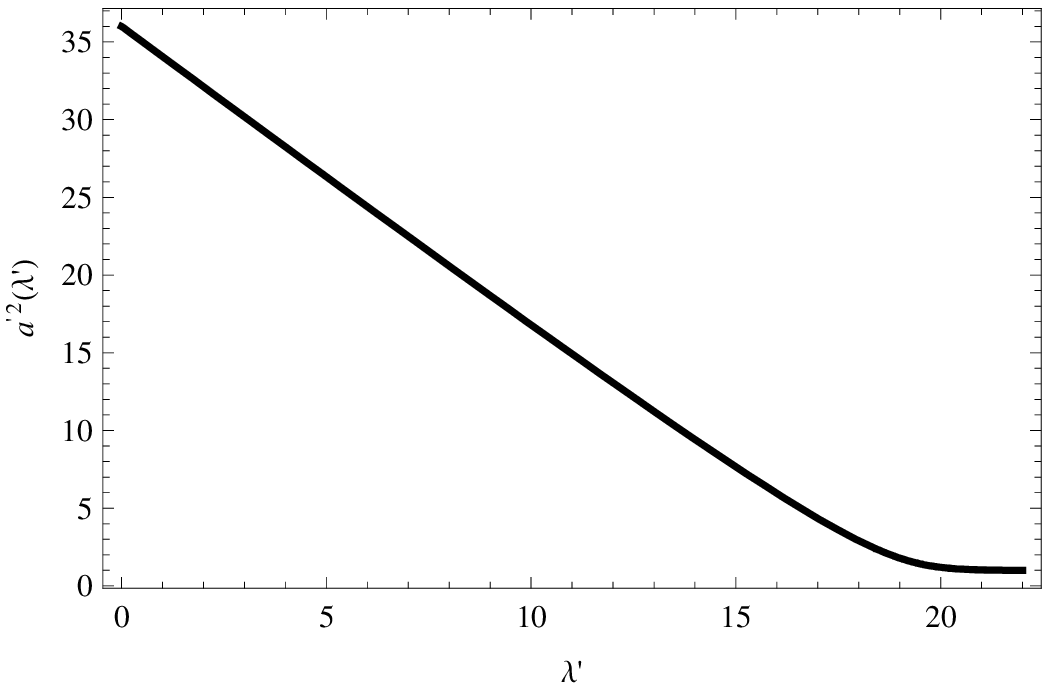} }
\subfigure[$a'(\lambda')^2<1,a'_0= 0.5$, $\alpha'<0$]{\includegraphics[width=0.32\textwidth]{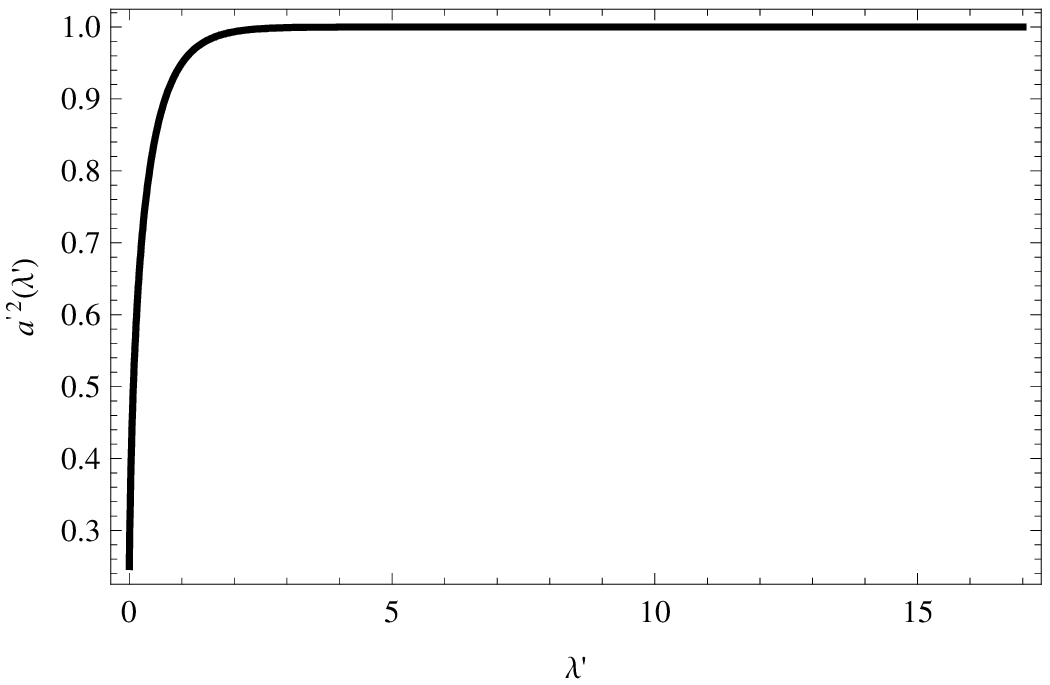} }
\caption{$a'^{2}(\lambda')$ vs $\lambda'$  for  $2nd$ order flow}
\label{2ndorderS2}
\end{figure}

The fact that for $\alpha'>0$ we get an ancient solution 
is shown in Fig.{\ref{2ndorderS2}} (top left). 
 If we consider $\alpha'<0$, which is essentially backward $2nd$ order flow 
upto  a scaling, we can see that the solution space is divided into two regions namely $a^2 <1$ and $a^2 >1$. It is also useful to note that  if we choose the initial condition $a_0=1$ for $\alpha'<0$ we get a soliton. 
The solution of the flow equations for  $a^2 >1$, $a^2 <1$ with $\alpha'<0$ are given as:
  \be\label{sol2ndS21}
 a'(\lambda')^{2}~+~\ln(a'(\lambda')^2-1)-a_0'^2~-~\ln(a_0'^{2}-1)=-2\lambda' ,~~~~~~~\mathrm{for} ~~a'^{2} (\lambda') >1
 \ee
 \be\label{sol2ndS22}
 a'(\lambda')^2~+~\ln(1- a'(\lambda')^2)-a_0'^2~-~\ln(1-  a_0'^2)=-2\lambda' ~~~~~~~\mathrm{for} ~~a'^{2} (\lambda') <1
\ee
The proof of the proposition is evident through the plots (in Fig.\ref{2ndorderS2}) 
and also from the 
general expressions given above. Here and henceforth, the values chosen
while obtaining the graphs are representative. We have also checked things
over the full, respective ranges.  
%\newpage
\subsubsection{\underline{3rd order flow on $S^2$}}
\begin{pro}\label{innpd}
Inclusion of the $3rd$ order terms for the flow on the 
canonical two-sphere, with any  choice of initial radius ($a'_0\ge 1$) 
and $\alpha' $ always gives rise to an ancient solution.
\end{pro}
The solution of the flow equation, taken up to $3rd$ order, 
will be:
\begin{equation}\label{sol3rds2}
%  \begin{split}
    a'(\lambda')^2~\mp~\f{1}{2}\ln(4a'(\lambda')^4\pm 4a'(\lambda')^2+5)
\mp~\f{3}{4}~ \tan^{-1} \f{(2a'(\lambda')^2\pm 1)}{2}=-2\lambda' + C
% \end{split}
 \end{equation}
 where $C=a_0'^2 \mp \f{1}{2}\ln(4a_0'^4 \pm 4a_0'^2+5) \mp \f{3}{4}~ \tan^{-1}
\f{(2a_0'^2 \pm 1)}{2}$. The  $\pm$ signs refer to $\alpha'>0$ and $\alpha'<0$ respectively.

\begin{figure}[htbp]
\centering
\subfigure[$a'_0= 1, \lambda'{_s}=0.0665$]{\includegraphics[width=0.4\textwidth]{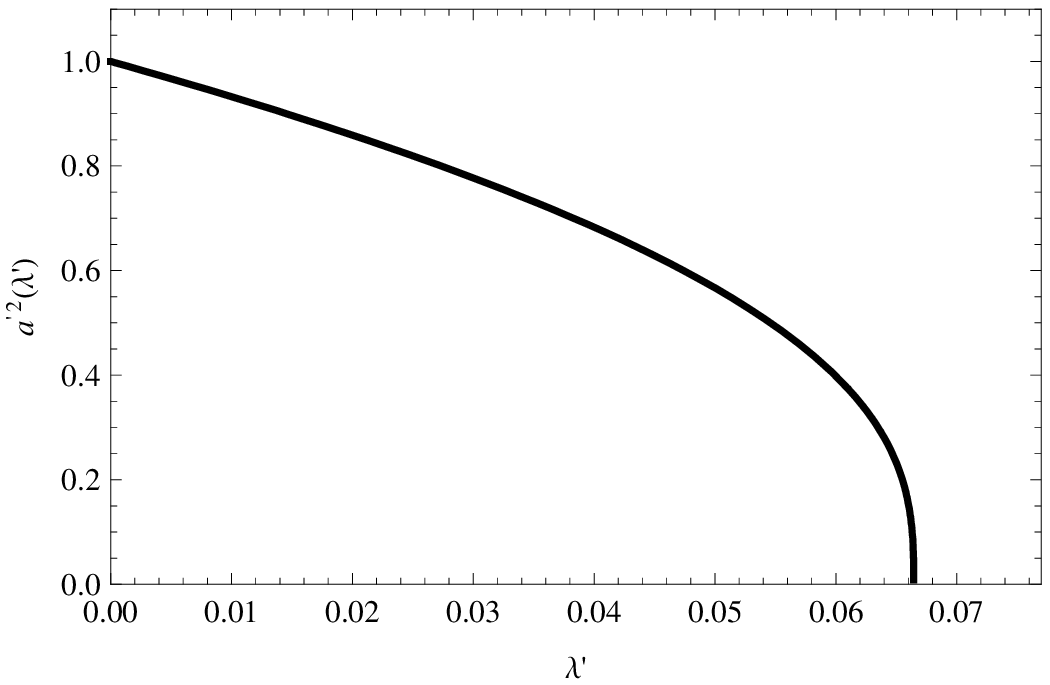} }
\subfigure[$a'_0= 1, \lambda'{_s}=0.152$]{\includegraphics[width=0.4\textwidth]{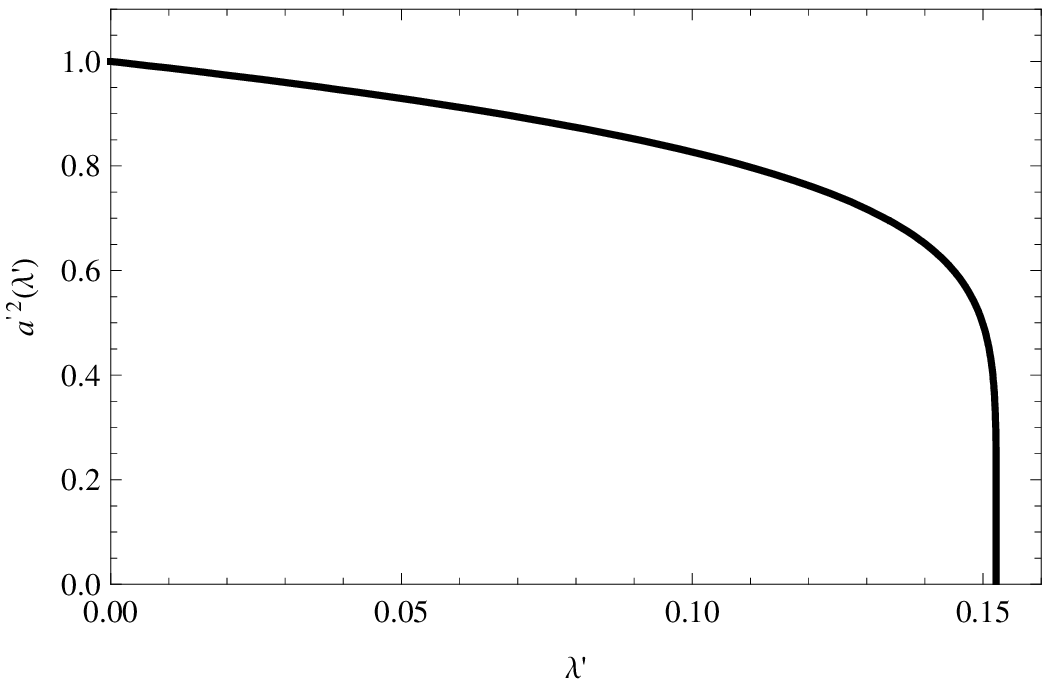} }
\subfigure[$a'_0= 2, \lambda'{_s}=0.288$]{\includegraphics[width=0.4\textwidth]{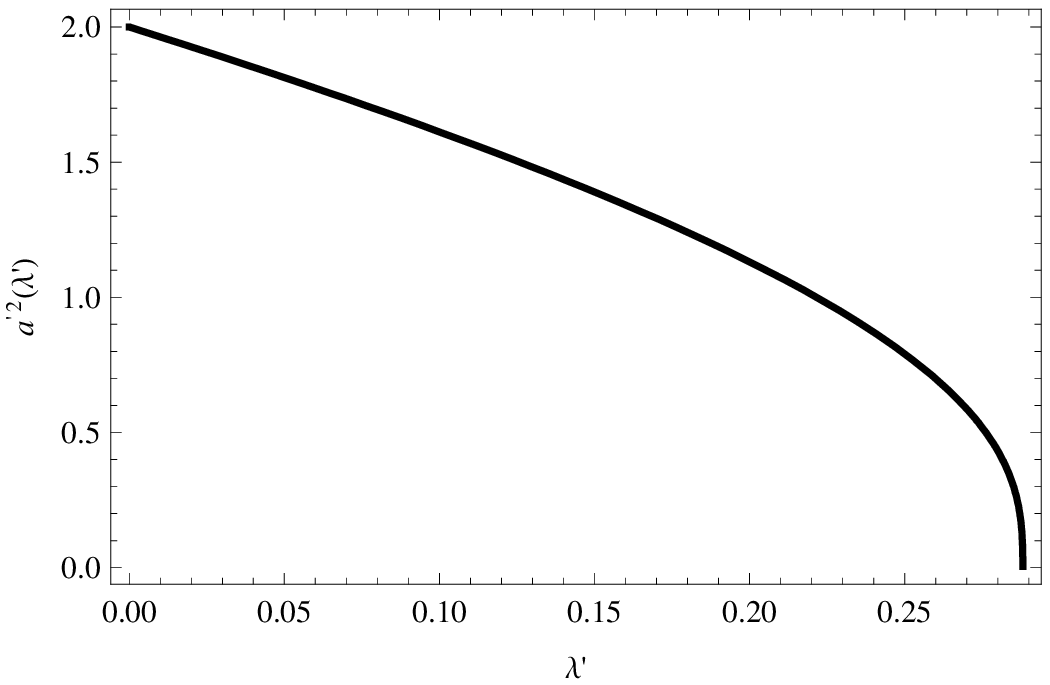} }
\subfigure[$a'_0= 2, \lambda'{_s}=0.696$]{\includegraphics[width=0.4\textwidth]{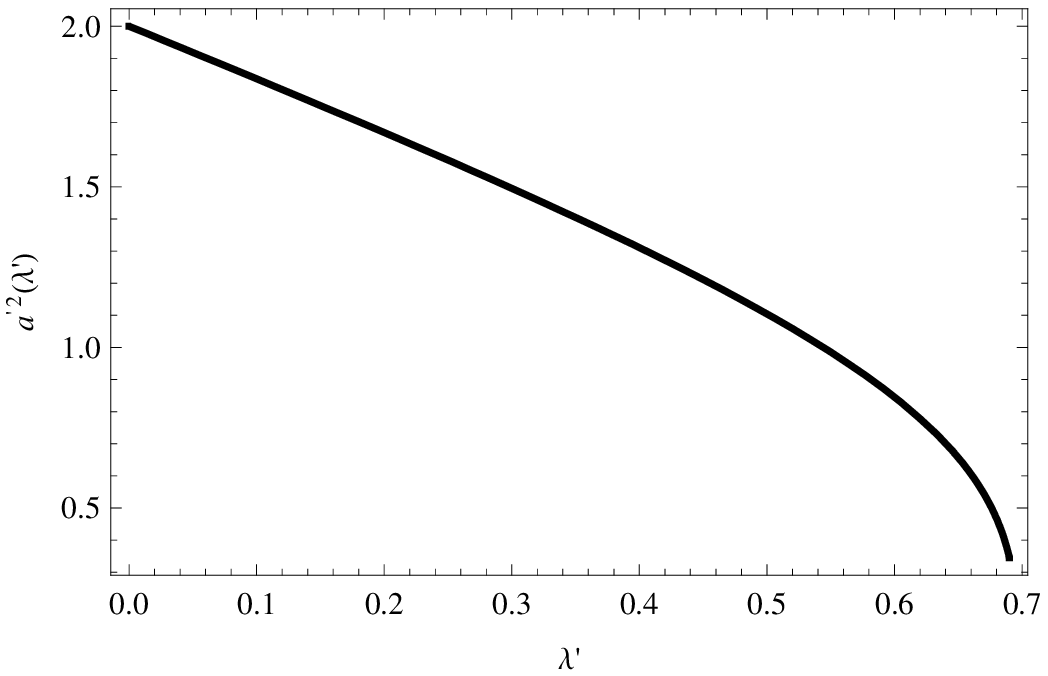} }
\caption{$a'^{2}(\lambda')$ vs $\lambda'$ for  $3rd$ order flow}
\label{3rdorderS2}
\end{figure}
From eqn.{\ref{sol3rds2}} and the figures the proposition can be proved 
without much difficulty.
Fig.\ref{3rdorderS2} illustrates the conclusions in the proposition. Figures on the left and right are for $\alpha'>0$ and $\alpha'<0$ respectively.
%\newpage
\subsubsection{\underline{4th order flow on $S^2$}}
\begin{pro}\label{innpd}
Inclusion of the $4th$ order terms in the flow on a canonical two-sphere,
we observe the following behavior:
\begin{itemize}
\item $ \alpha'>0$ and any choice of  $a'_{0}$ gives rise to ancient solution.	
\item $ \alpha'<0,a'_{0}=1.3048$: soliton. For $a'_{0}\neq1.3048$ we find,
$a'_{\infty}$=const
\item $ \alpha'<0;a'_{0}<1.3048$: immortal solution and . $a'_{0}>1.3048$: eternal solution.
		
\end{itemize}
\end{pro}
The solution of the flow equation with the fourth order term included
is found to be,
\be\label{}
a'^{2}+c_1\ln[a'^{2}-\eta ]-\frac{c_2}{2}\ln\left[a'^{4}+\beta  a'^{2}+ \delta \right]-\frac{2c_3-\beta c_2}{2\sqrt{\delta -\frac{\beta ^2}{4}}}tan^{-1}\left[\frac{a'^{2} + \frac{\beta }{2}}{\sqrt{\delta -\frac{\beta ^2}{4}}}\right]=-2\lambda'
+ C
\ee
where the constants are:
\be \label{4thorderconst1}
C=a_0'^{2}+c_1\ln[a_0'^{2}-\eta ]-\frac{c_2}{2}\ln\left[a_0'^{4}+\beta  a_0'^{2}+ \delta \right]-\frac{2c_3-\beta c_2}{2\sqrt{\delta -\frac{\beta ^2}{4}}}tan^{-1}\left[\frac{a_0'^{2} + \frac{\beta }{2}}{\sqrt{\delta -\frac{\beta ^2}{4}}}\right]
\ee
\begin{subequations}\label{4thorderconst2}
		\begin{align}
		c_1\approx 0.7544\hspace{10pt} & c_2\approx -0.2456\hspace{10pt}  c_3\approx 1..3618\label{}\\
		\eta \approx 1.7024\hspace{10pt} & \beta \approx 0.7024\hspace{10pt}  \delta \approx 2.4457\label{}
		\end{align}
	\end{subequations}
 \begin{figure}[htbp] 
\centering
\subfigure[$a'_0= 1, \lambda'{_s}=0.019$]{\label{4thS2one}\includegraphics[width=0.4\textwidth]{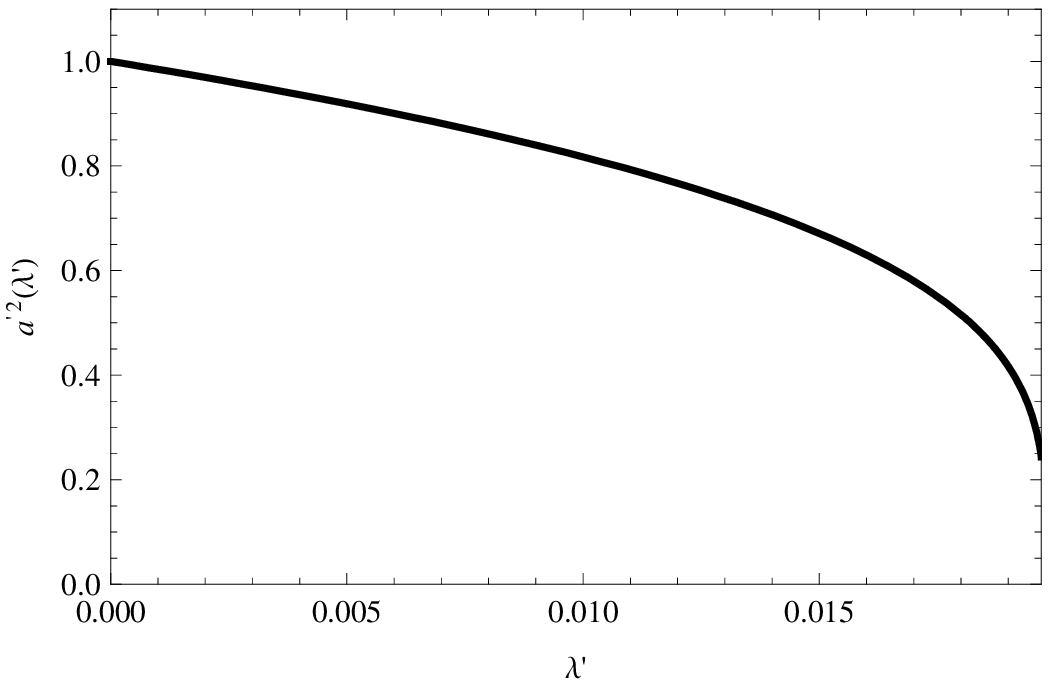} }
\subfigure[$a'_0= 1, \lambda'{_s}=-0.0211$]{\label{4thS2two}\includegraphics[width=0.4\textwidth]{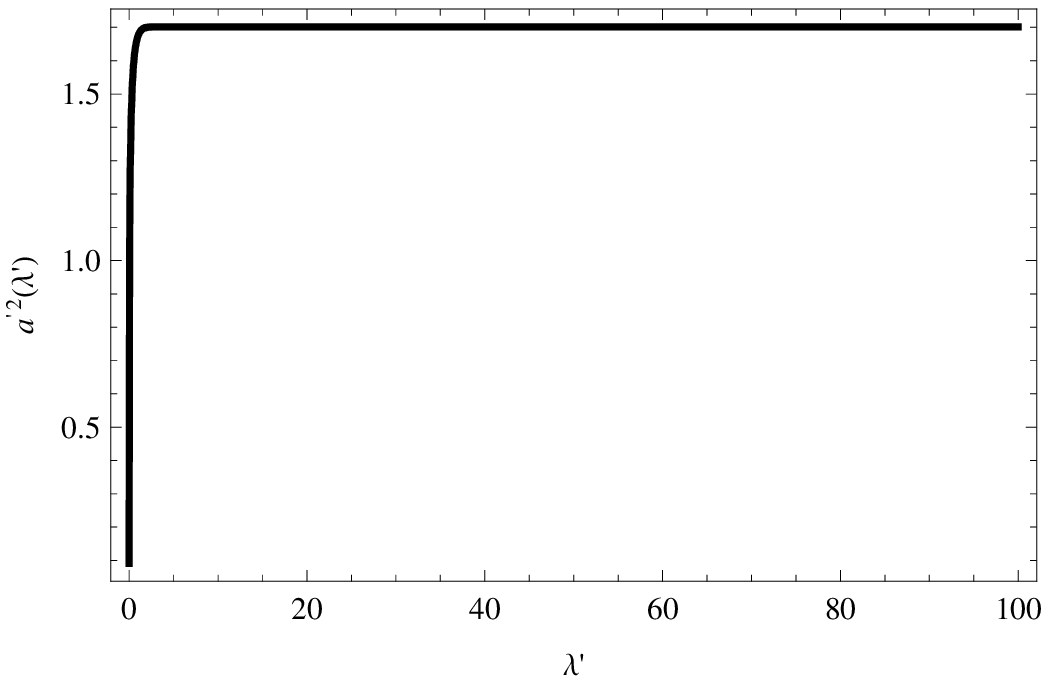} }\\
\subfigure[$a'_0= 3$]{\label{4thS2three}\includegraphics[width=0.4\textwidth]{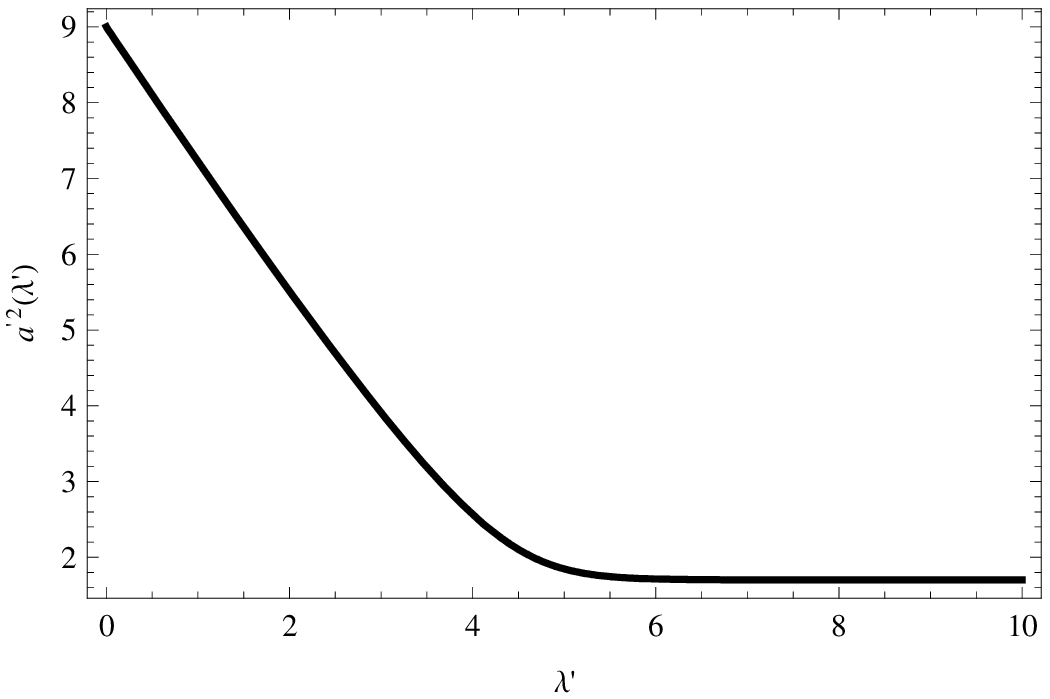} }
\caption{$a'^{2}(\lambda')$ vs $\lambda'$ for   $4th$ order flow}
\label{4thorderS2}
\end{figure}
Fig.\ref{4thS2one} shows the ancient solution, while Fig.\ref{4thS2two} 
illustrates the ancient solution 
for $a_0'<1.3048(a'_{\infty}=constant)$. Fig.\ref{4thS2three} 
with $a_0'>1.3048(a'_{\infty}=constant)$ demonstrates the eternal solution. 
%\newpage
\subsubsection{\underline{2nd order flow on $H^2$}}
 \begin{pro}\label{innpd}
 $2nd$ order flow on hyperbolic space with  $\alpha'>0$ generates two kinds of final metrics depending on the initial scale factors. For $b'(\lambda')^2>1$ it is expanding and for $b'(\lambda')^2<1$ it is converging. In both cases the scale factor asymptotically tends to $1$ in backward time( $b'_{-\infty}=1$).  For $\alpha'<0$,  we  obtain    an immortal solution. 
\end{pro}
 
\begin{figure}[htbp] 
\centering
%\subfigure[$a_0= 1, T_s=0.153$]{\includegraphics[width=0.4\textwidth]{2ndS2} }
%\subfigure[$a_0= 1$, soliton]{\includegraphics[width=0.4\textwidth]{2ndS2neg} }
\subfigure[$b'(\lambda')^2>1,b'_0= 8$, $\alpha'>0$]{\includegraphics[width=0.4\textwidth]{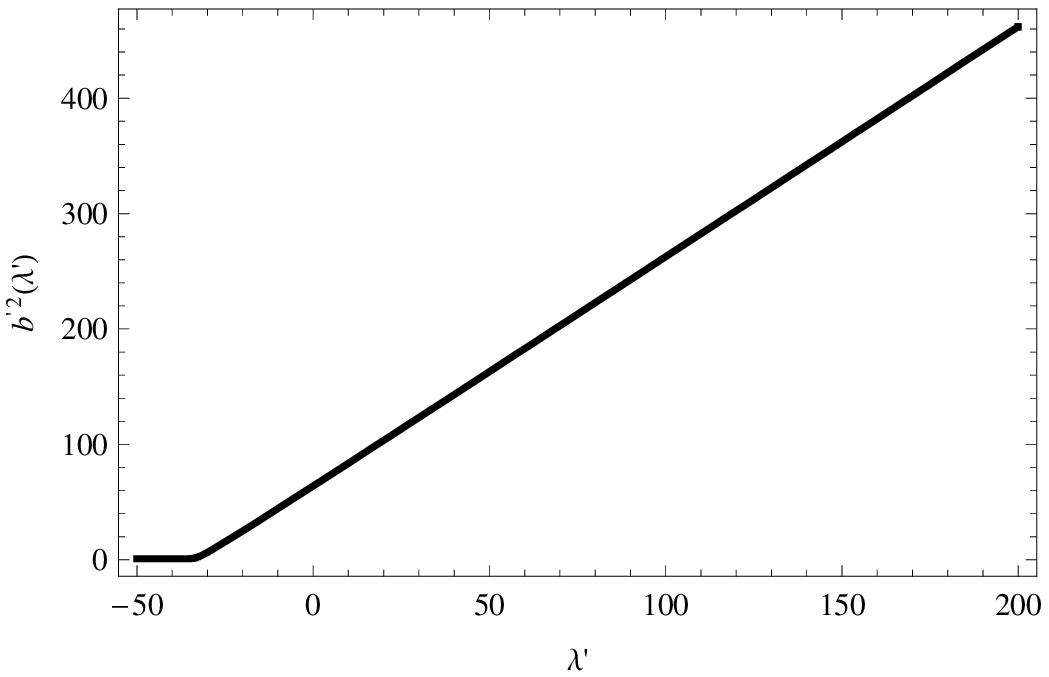} }
\subfigure[$b(\lambda)^2<1,b'_0= 0.5$, $\alpha'>0$]{\includegraphics[width=0.4\textwidth]{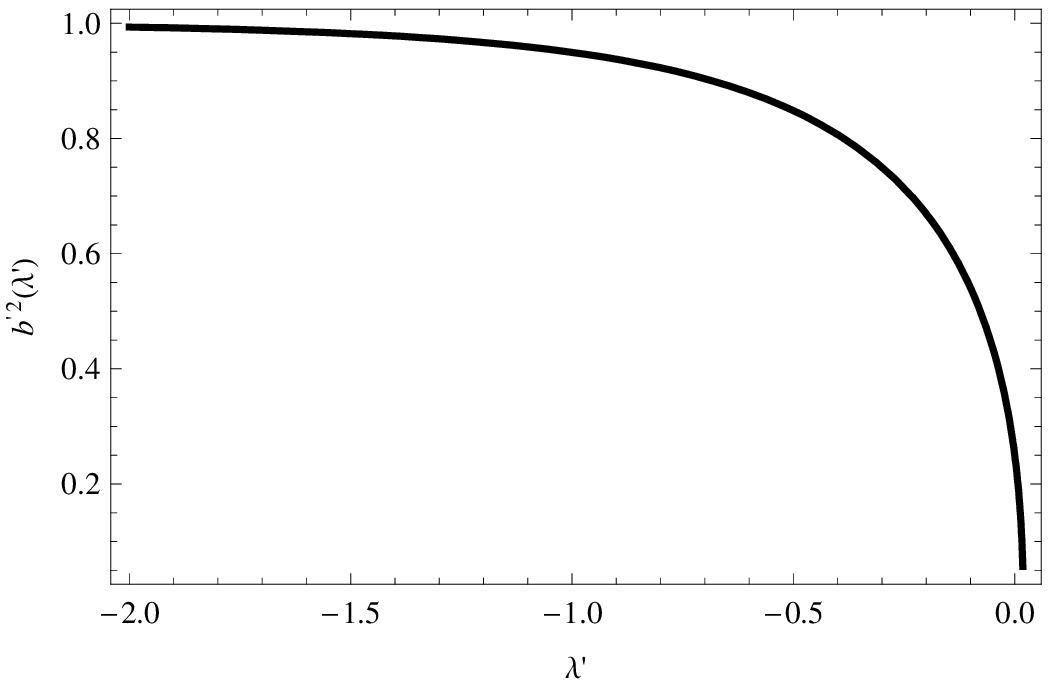} }
\subfigure[$b'_0= 1$, $\alpha'>0$,soliton]{\includegraphics[width=0.4\textwidth]{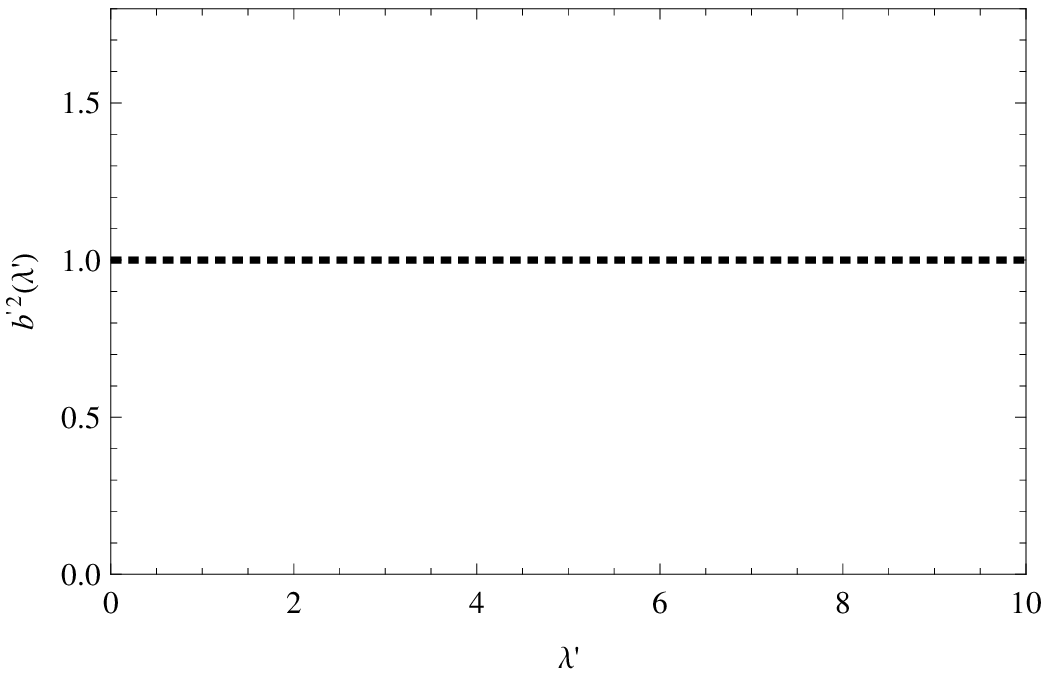} }
\subfigure[$b'_0= 0.4$, $\alpha'<0$]{\includegraphics[width=0.4\textwidth]{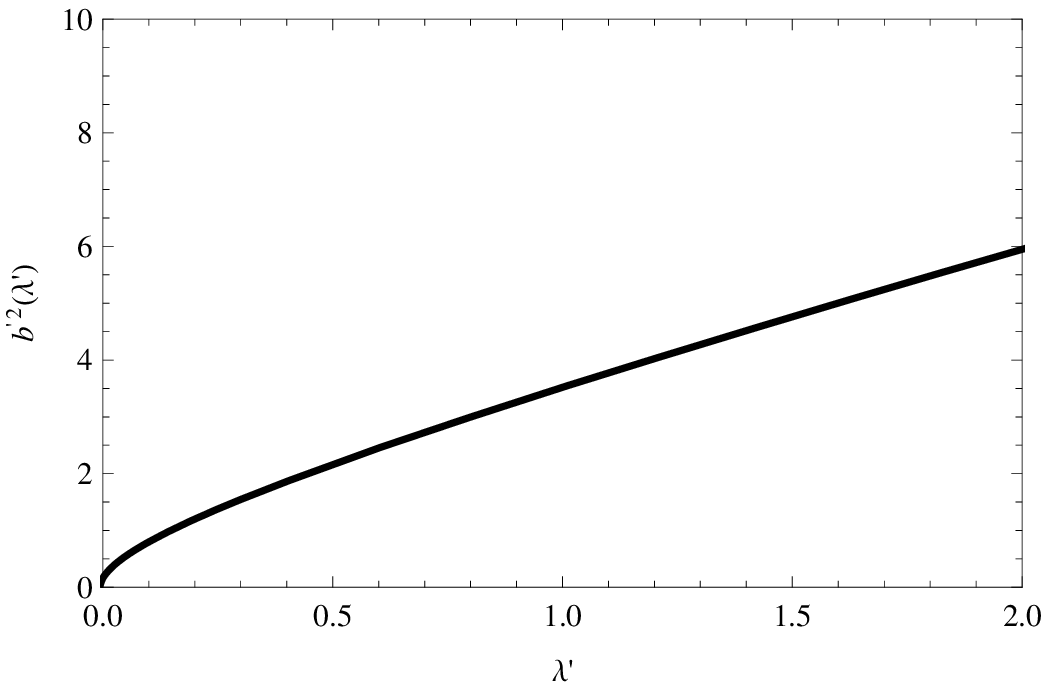} }
\caption{$b^{'2}(\lambda')$ vs $\lambda'$ for   $2nd$ order flow}
\label{2ndorderH2}
\end{figure}

When $\alpha'>0$, the flow equation $(k=-1)$, has two solutions for 
$b'^2(\lambda')>1$ and $b'^2(\lambda')<1$. These are given as: 
 \be\label{sol2ndH21}
 b'(\lambda')^2~+~\ln(b'(\lambda')^2-1)-b_0'^2~-~\ln(b_0'^2-1)=2\lambda',  ~~~~~~~ \mathrm{for} ~~b'^{2} (\lambda') >1
 \ee
 \be\label{sol2ndH22}
 b'(\lambda')^2~+~\ln(1- b'(\lambda')^2)-b_0'^2~-~\ln(1-  b_0'^2)=2\lambda' ~~~~~~~ \mathrm{for} ~~b'^{2} (\lambda') <1
\ee
It can be seen that $b_0 =1$ corresponds to a soliton. 
If we consider $\alpha'<0$, which essentially produces an immortal solution, 
we can see that it expands along the flow. The conclusions mentioned in the 
above proposition are demionstrated in Fig.{\ref{2ndorderH2}}   

%\newpage
 \subsubsection{\underline{3rd order flow on $H^2$}}
\begin{figure}[htbp]
\centering
\subfigure[$b_0'= 0.4$, $\alpha'>0$]{\includegraphics[width=0.4\textwidth]{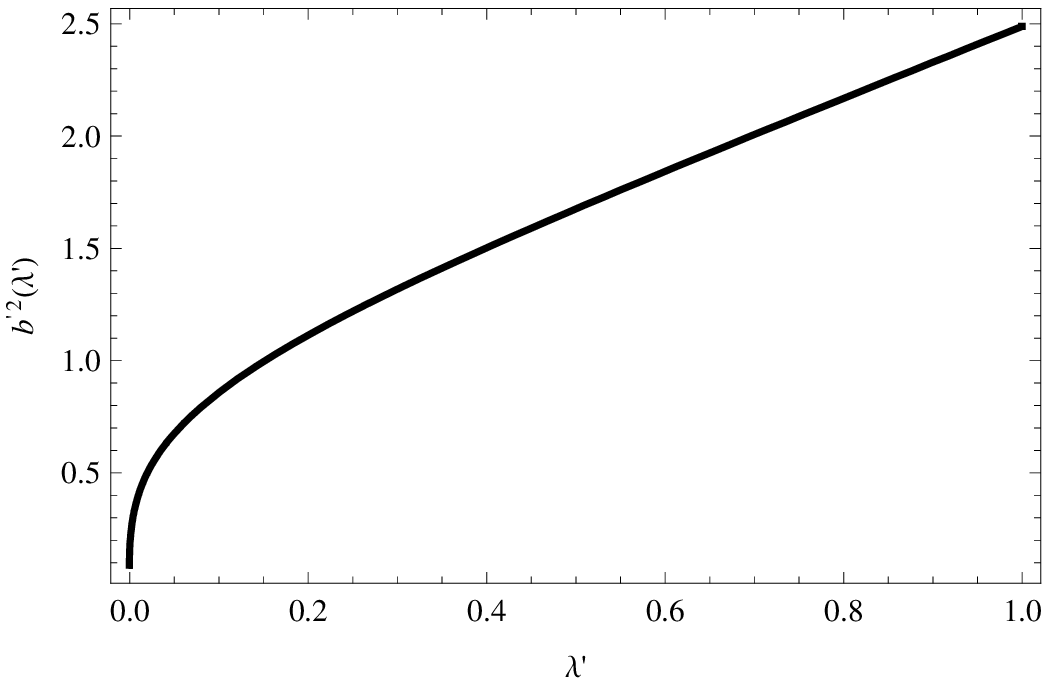}}
\subfigure[$b_0'= 0.4$,
$\alpha'<0$]{\includegraphics[width=0.4\textwidth]{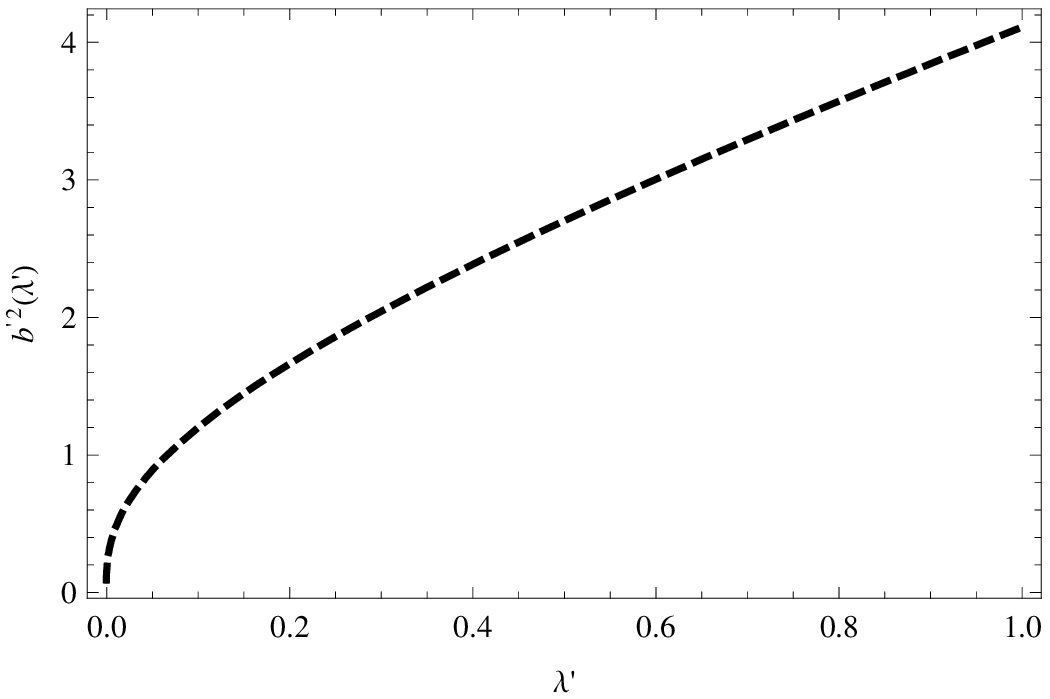} }
\caption{$b^{'2}(\lambda')$ vs $\lambda'$ for   $3rd$ order flow}
\label{3rdorderH2}
\end{figure}

  \begin{pro}\label{pp3rdh2}
Inclusion of the $3rd$ order terms in the flow on hyperbolic space, 
with any choice of initial radius ($b_0'\ge 1$) and any choice of  
$\alpha'$ gives rise to an immortal solution.
\end{pro}
   For $\alpha'>0$ the solution turns  out to be:
 \begin{equation}\label{sol3rdh2}
  \begin{split}
 b'(\lambda')^2~\pm \f{1}{2}~\ln(4b'(\lambda')^4 \mp
4b'(\lambda')^2+5)-~\4{3}{4}~ tan^{-1}(\f{2b'(\lambda')^2 \mp1}{2})\\
    -b_0'^2~\mp~\f{1}{2}~\ln(2b_0'^4\mp2b_0'^2+5)+~\4{3}{4}~
tan^{-1}(\f{2b_0'^2\mp1}{2})=2\lambda'
 \end{split}
 \end{equation}
The conclusion in Prop. {\ref{pp3rdh2}} is shown in Fig.{\ref{3rdorderH2}}

\subsubsection{\underline{4th order flow on $\mathbb{H}^2$}}
  \begin{pro}\label{pp4thorderH2}
Inclusion of the $4th$ order term in the flow on hyperbolic space, 
yields two cases :\\
1a) for $\alpha'>0$ and $b^{'2}_{0}\leq 1$ (except 1.3048, which corresponds to a soliton)  we have an ancient solution and $b'_{-\infty}=const.$\\
1b) for $\alpha'>0$ and $b^{'2}_{0}> 1$ we have an eternal solution and 
$b^{'}_{-\infty}=const.$\\
2) for $\alpha'<0$ and any choice of $b_0'$ we have an expanding immortal solution.\\
 \end{pro}
 The solution of this equation would be:\\
\be\label{}
b'^{2}\pm c_1\ln[b'^{2}\mp \eta ]\mp \frac{c_2}{2}\ln\left[b'^{4}\pm \beta  a'^{2}+ \delta \right]-\frac{2c_3-\beta c_2}{2\sqrt{\delta -\frac{\beta ^2}{4}}}tan^{-1}\left[\frac{b'^{2} \pm \frac{\beta }{2}}{\sqrt{\delta -\frac{\beta ^2}{4}}}\right]=2\lambda'+ C
\ee
where the constants are given in Eq.\ref{4thorderconst1} and Eq.\ref{4thorderconst2}. The upper and lower sign corresponds to the solution for $\alpha'>0$ and $\alpha'<0$ respectively.

\begin{figure}[htbp] 
\centering
\subfigure[$b_0'= 2$, $\alpha'>0$]{\includegraphics[width=0.30\textwidth]{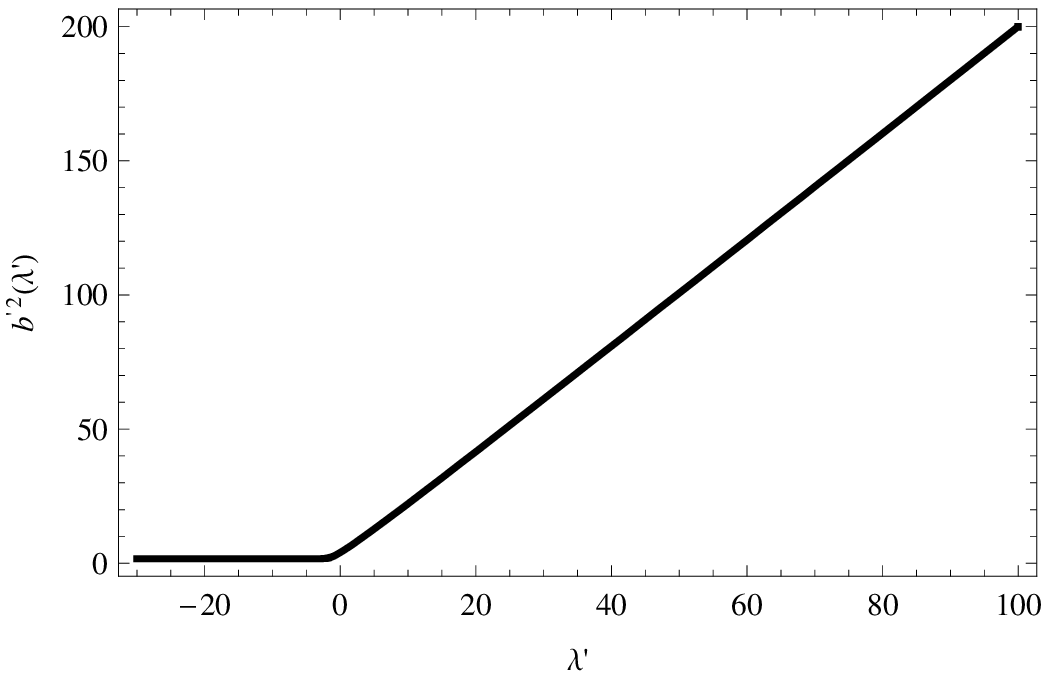} }
\subfigure[$b_0'= 1$, $\alpha'>0$]{\includegraphics[width=0.30\textwidth]{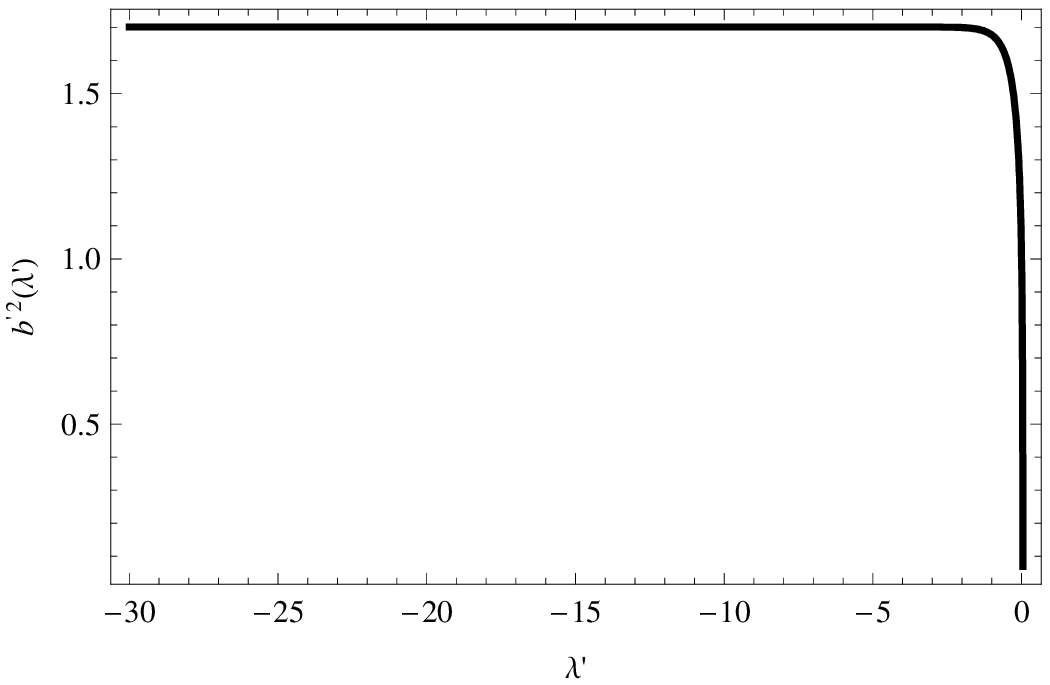} }\\
\subfigure[$b_0'= 1$, $\alpha'<0$]{\includegraphics[width=0.30\textwidth]{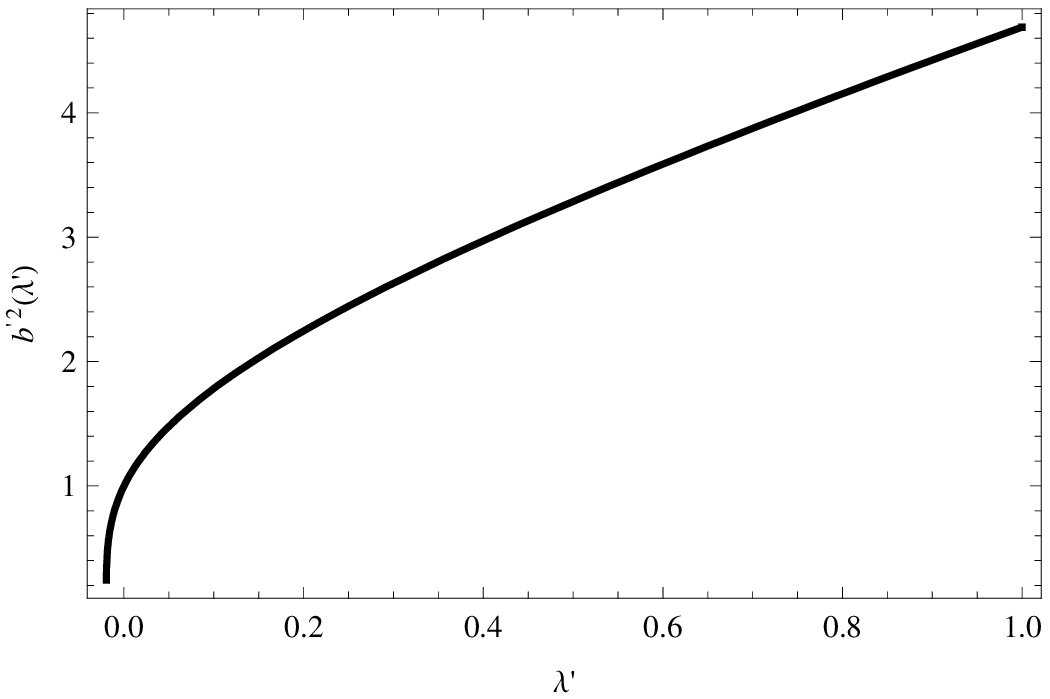} }
\caption{$b'^{2}(\lambda')$ vs $\lambda'$ for   $4th$ order flow}
\label{4thorderH2}
\end{figure}

 All the statements in proposition{\ref{pp4thorderH2}} can be checked, as before, by solving the flow equations. The graphs in Fig.{\ref{4thorderH2}} illustrate these facts clearly.
Table{\ref{table1}} provides a comparison of the results for $\mathbb{S}^2$ and $\mathbb{H}^2$ at various orders.
%\newpage
\begin{table}[htbp]
\caption{Comparison of flows on  $\mathbb{S}^2$ and $\mathbb{H}^2$ at $2$nd order and $4$th order}
\begin{center}
\begin{tabular}{|c||c|c||}
\hline
$$ & $\mathbb{S}^{2}(K>1)$ & $\mathbb{H}^{2}(K<1)$ \\
\hline\hline
$(2nd)$                                & \,\,\,\,\,\,$a_0' > 0$,\textrm{ancient soln. }~~~~~~~~~                                             & $b_0'=1$, \textrm{soliton}\\
     $\alpha'>0$                    &\textrm{ shrinking }            ~~~~~~~~~                                & $b_0'^2<1$, $b'_{-\infty}=1$,\textrm{converging}\\
$                $                          & $              $             ~~~~~~~~~                                & $b_0'^2>1$, $b'_{-\infty}=1$,\textrm{expanding}\\
\hline
$(4th)$                                & $a'_0>0$,  \textrm{ancient soln.}                      &  $b'_{0}\leq 1$, \textrm{ancient }, $b'_{-\infty}=const.$ \\
    $\alpha'>0$                   & \textrm{shrinking}                                                             & $b'_{0}> 1, eternal,expanding$\\
                                             &                                                                                              & $b'_{0}=1.3048$, soliton\\

\hline
$(2nd)$                                & $a'_0=1$,  \textrm{soliton}                                         &  $$\\
$\alpha'<0$                        & $a_0'^2<1$, $a'_{\infty}=1$                                 & $b_0'>0$ \textrm{immortal,expanding}\\
$                 $                         & $a_0'^2>1$, $a'_{\infty}=1$                                        & $$\\

\hline
$(4th)$                                & $a'_0=1.3048$,  \textrm{soliton}                                &  any $b'_{0}$, \textrm{immortal soln.}  \\
$\alpha'<0$                        & $a_0'^2>1$, $a'_{\infty}=const.$                               & \textrm{expanding}\\
$                 $                         & $a_0'^2<1$, $a'_{\infty}=const.$                              &                                   \\
\hline\hline

\end{tabular}
\end{center}
\label{table1}
\end{table}
%\newpage
%%%%%%%%%%
\subsubsection{\underline{Higher order flows on $Ag_{can}^{\mathbb S^2}~\oplus~Bg_{can}^{N}$}}\
We now briefly discuss flows on an unwarped product manifold before moving on to
warped products in the next section.

\begin{figure}[htbp]
\centering
\subfigure[First][$(A_0,B_0)= (5,3),\alpha'>0,\lambda'_{s}=1.6041$]{\label{firstfig}\includegraphics[width=0.39\textwidth]{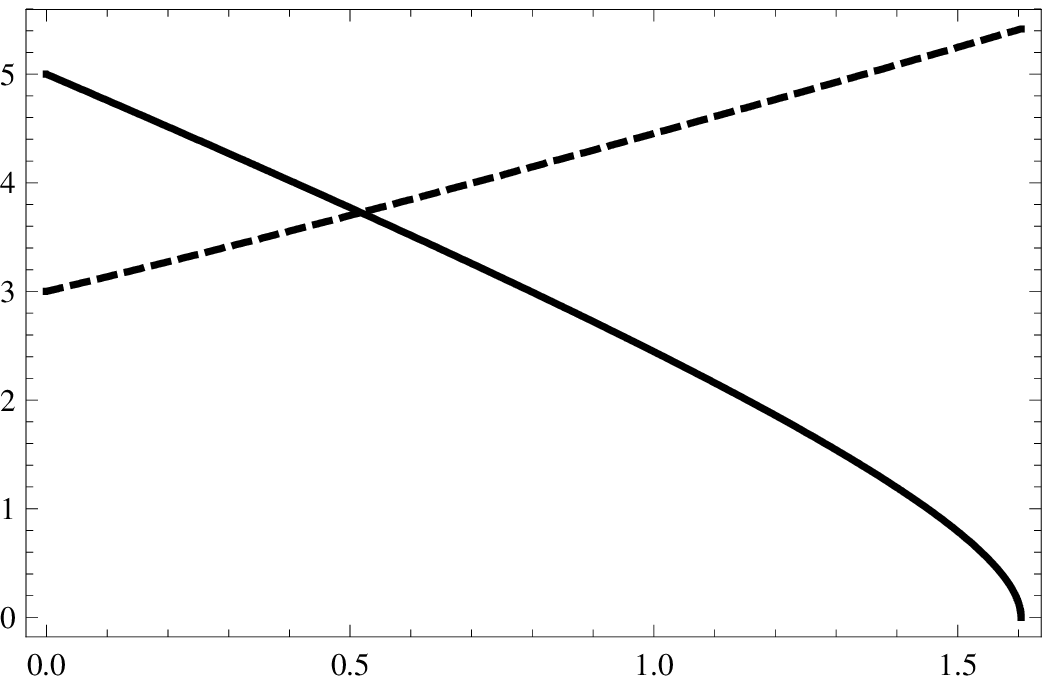}}
\subfigure[Second][$(A_0,B_0)= (5,3),\alpha'<0$ ]{\label{2ndfig}\includegraphics[width=0.4\textwidth]{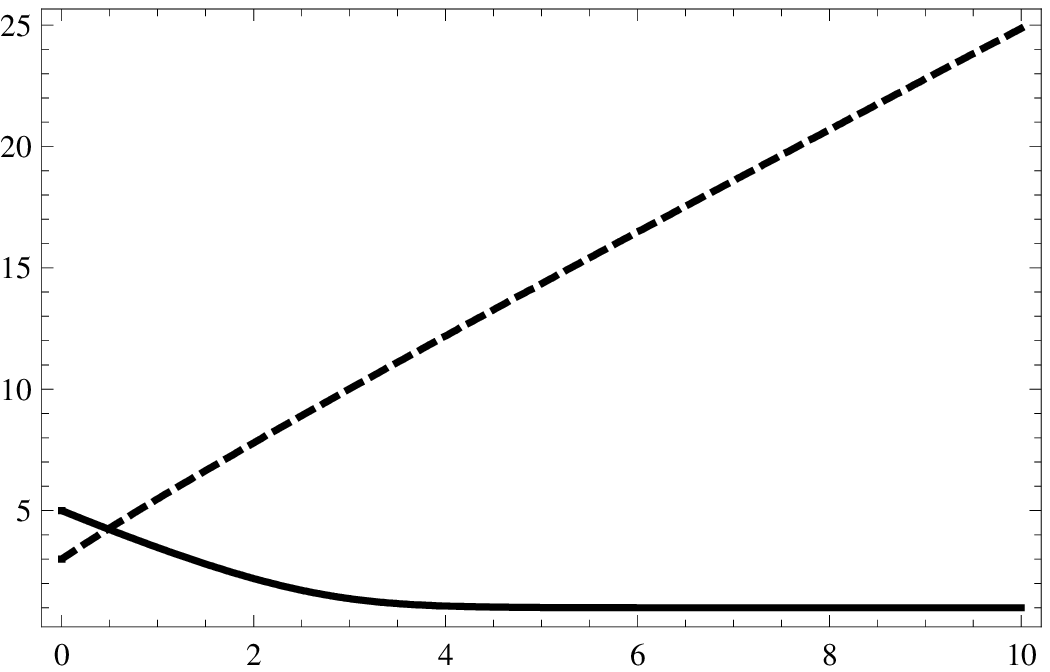} }
\subfigure[Third][$(A_0,B_0)= (5,3),\alpha'>0$ ]{\label{3rdfig}\includegraphics[width=0.4\textwidth]{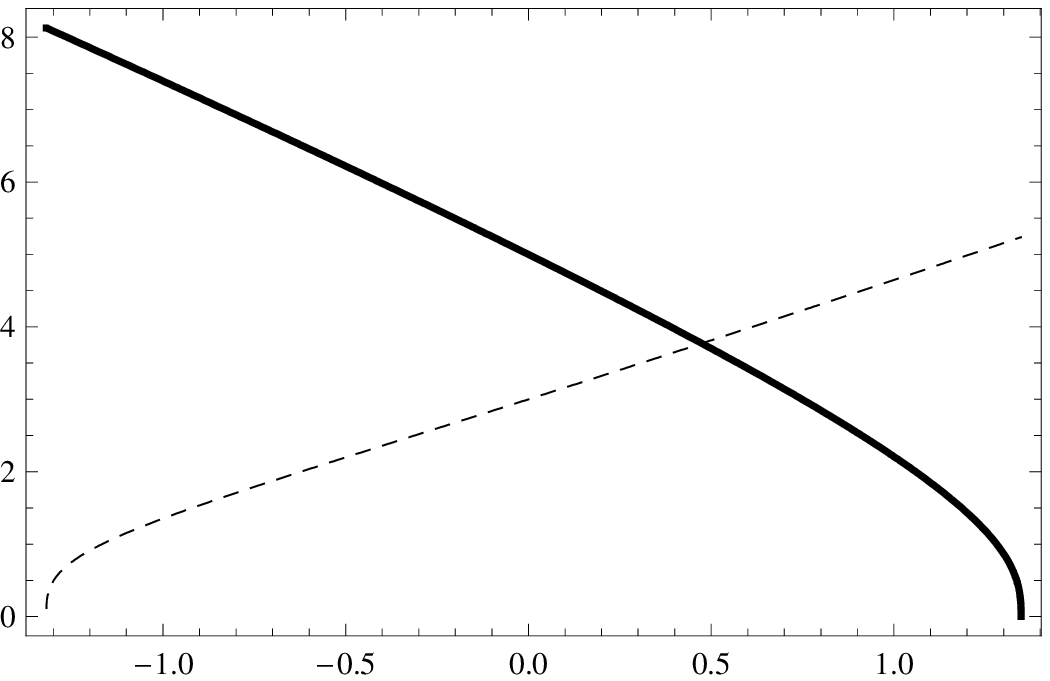}}
\subfigure[Forth][$(A_0,B_0)= (5,3),\alpha'<0$ ]{\label{4thfig}\includegraphics[width=0.4\textwidth]{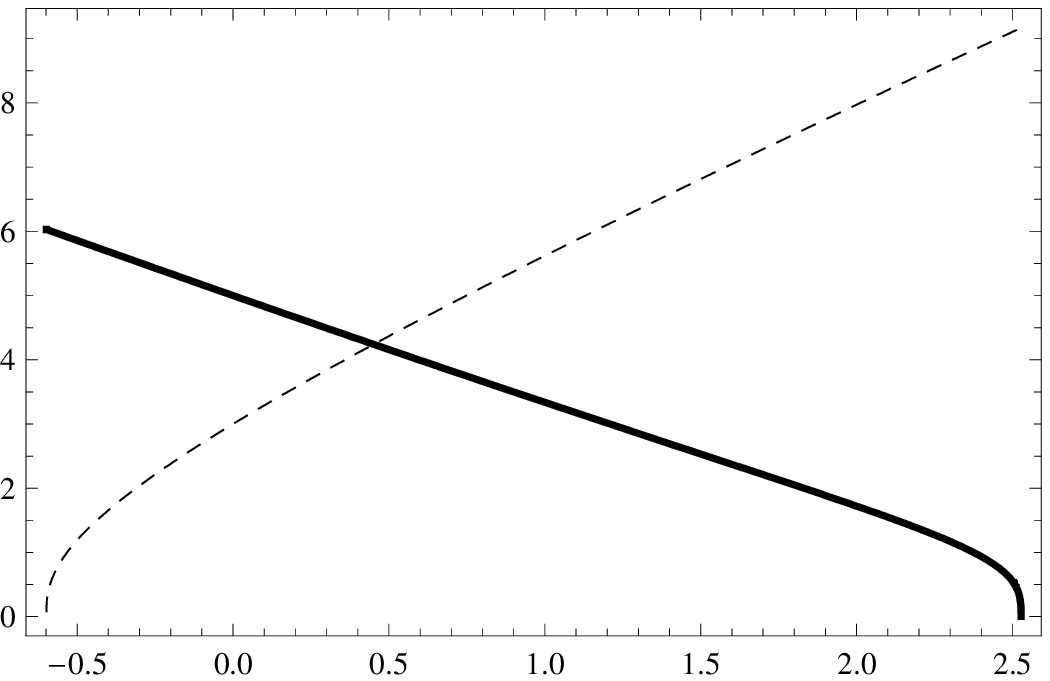} }
\caption{($A(\lambda')$,$B(\lambda')$) vs $\lambda$ for $2nd$ order flow ((a),(b))and $3rd$ order flow((c),(d))}
\label{2ndorderS2XN}
\end{figure}

Let N be a manifold(surface) of genus $g\ge 2$ and $g_{can}^N$ is a constant negative curvature($-1$).
 From Eqn.{\ref{gensolsurf1}} we can write the flow equations up to $4th$ order
(note that they are decoupled equations because the product is unwarped):
\ba\label{S2XNgeneqn}
\f{dA}{d\lambda'}=-2~-~\alpha'\f{2}{A}~-~\alpha'^2\f{5}{2}\f{1}{A^2}~-~2\alpha'^3(\4{59}{24}\zeta(3)+\4{29}{24})\4{1}{A^3}
\\
\f{dB}{d\lambda'}=2~-~\alpha'\f{2}{B}~+~\alpha'^2\f{5}{2}\f{1}{B^2}~-~2\alpha'^3(\4{59}{24}\zeta(3)+\4{29}{24})\4{1}{B^3}
 \ea
In this case, for $2nd$  and $4th$ order flow we can have ancient as well as immortal solution depending on $\alpha'$ ($>0$ or $<0$), respectively. 
Here Fig.\ref{firstfig} and Fig.\ref{2ndfig} show such examples (for $2nd$ order), 
of ancient and immortal solutions, respectively. We can also see from 
Fig.{\ref{2ndfig}} that the scale factor corresponding to $\mathbb{S}^2$ approaches a constant for $\alpha'<0$. The $3rd$ order flow is neither ancient nor immortal 
but maintains the same profile with the rest of its class.
This may be seen in Fig.{\ref{3rdfig}} and Fig.{\ref{4thfig}} for both choices of $\alpha'$. 
Here the continuous  and dashed lines represent the solutions for 
$A(\lambda')$ and $B(\lambda')$ respectively.
\subsection{Perturbative RG flow domain}
\begin{figure}[htbp]
\centering
\subfigure[$\alpha' > 0$]{\includegraphics[width=0.4\textwidth]{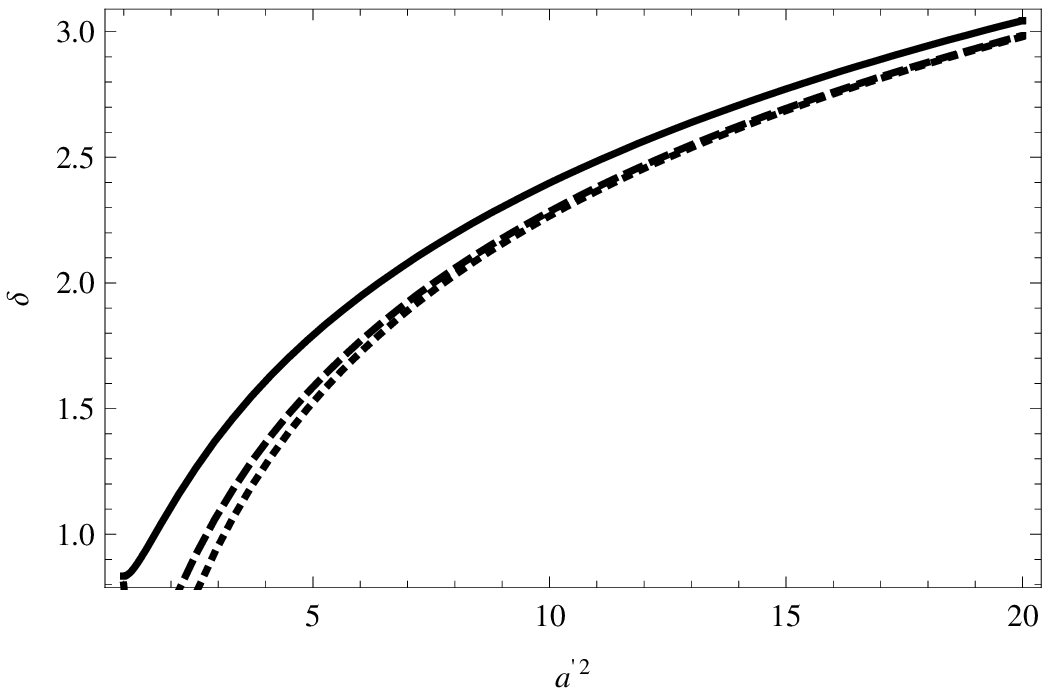}\label{devsph>0}}
\subfigure[$\alpha' < 0$]{\includegraphics[width=0.4\textwidth]{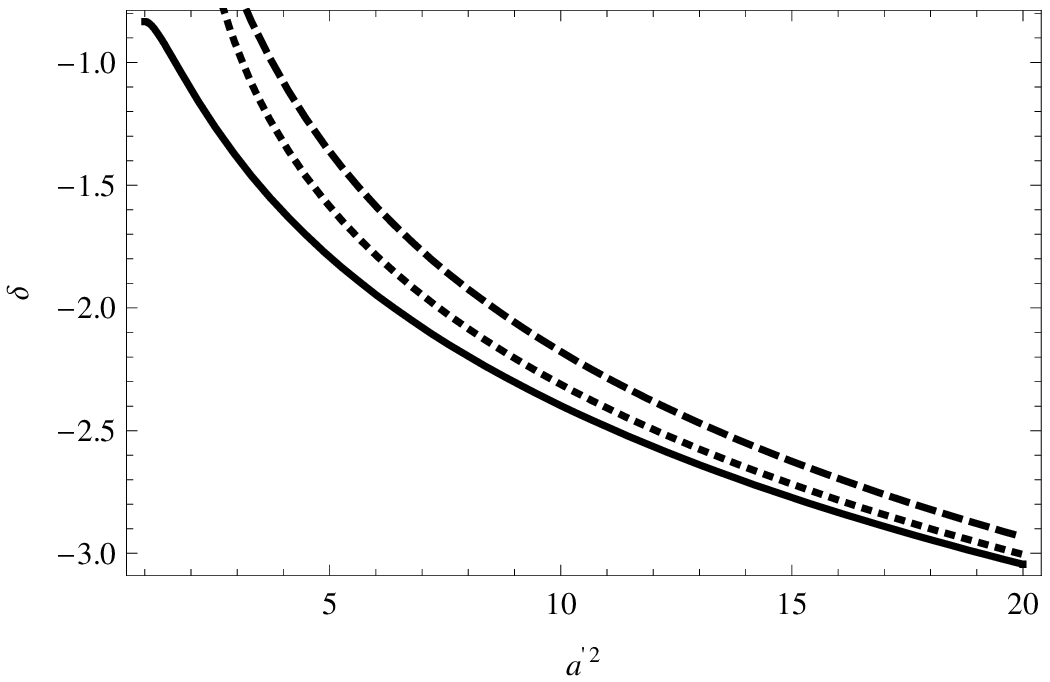}\label{devsph<0}}
\caption{Spherical surface:The plots for  $\delta$ at different orders, $2^{nd}$ (continuous line), $3^{rd}$ (dashed line),$4^{th}$ (dotted line).}
\label{dev expansionsphere}
\end{figure}

\begin{figure}[htbp]
\centering
\subfigure[$\alpha' >
0$]{\includegraphics[width=0.4\textwidth]{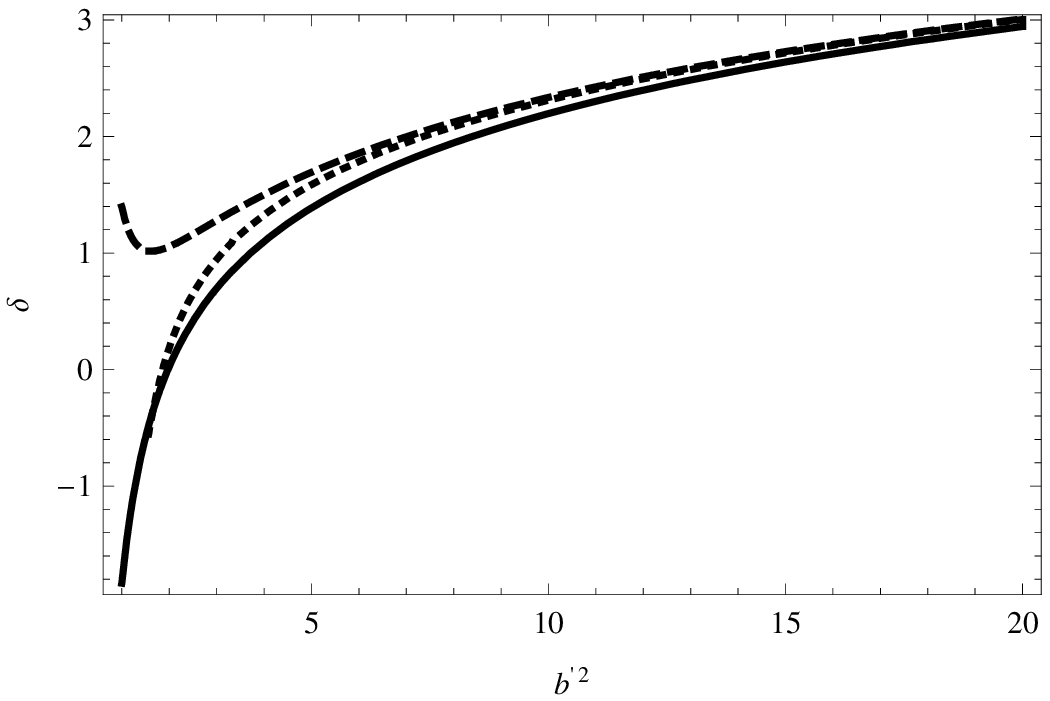}\label{devhyp>0}}
\subfigure[$\alpha' <
0$]{\includegraphics[width=0.4\textwidth]{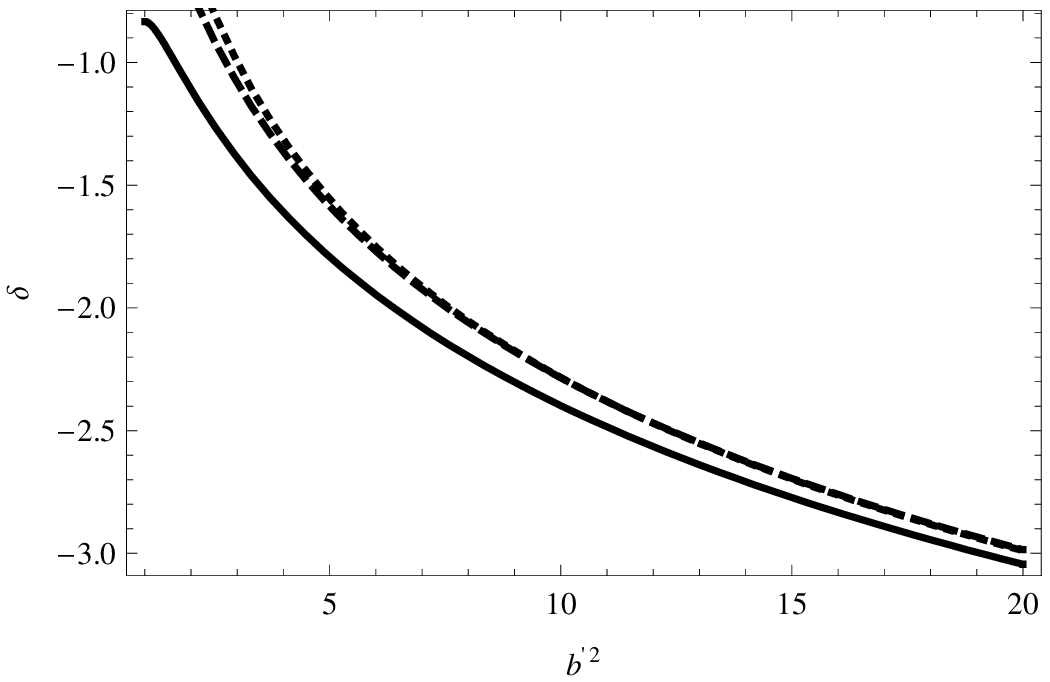}\label{devhyp<0}}
\caption{Hyperbolic surface:The plots for  $\delta$ at different orders, $2^{nd}$
(continuous line), $3^{rd}$ (dashed line),$4^{th}$ (dotted line).}
\label{dev expansionhyperbolic}
\end{figure}

Finally, we look at the case where $\4{1}{x'^{2}}\ll1$ or $x'^{2}\gg1$. 
In this regime, the RG Flow assumption of small curvatures is valid. 
We look at the behavior 
of the flow equations(Eq.\ref{gensolsurf1} and Eq.\ref{gensolsurf2}) at 
various orders, for spherical and hyperbolic surfaces. To obtain the
correction at various orders, we  
expand the solutions  at various orders and try to understand the 
characteristics of the RG flow within the abovementioned limit. 
To achieve our goals we consider a power series 
in $\4{1}{x'^{2}}$. Below, we show the expansion for $\alpha'>0$ in the case 
of a spherical surface at $2nd$ and $3rd$ order respectively. \\
	\begin{subequations}\label{expsphpos}
		\begin{align}
			-2\lambda' + C_2 & =a'^{2} + \ln(\4{1}{a'^{2}}) - \4{1}{a'^{2}} + \4{1}{2a'^{4}} - \4{1}{3a'^{6}}+ \ldots \\
			-2\lambda' + \tilde C_3 & =  a'^{2} + \ln(\4{1}{a'^{2}}) + \4{1}{4a'^{2}} - \4{3}{4a'^{4}} + \4{19}{48a'^{6}}+ \ldots 
					\end{align}
	\end{subequations}
The constant terms are $C_2=\ln(a_0'^{2}-1)-a_0'^2$ and $\tilde C_3=C+\4{3\pi}{8}+\4{\ln4}{2}$ where $C$ comes from Eq.{\ref{sol3rds2}}. The expansions 
for $\alpha'<0$ (spherical surface) and hyperbolic surface 
(both $\alpha'<0$ and $\alpha'>0$) can be obtained in a similar fashion though
the $\alpha'<0$ case is not important in the $\sigma$-model context. \\
Eq.{\ref{expsphpos}} can be written as $-2\lambda' + C = a'^{2} + \delta$ where $\delta$ represents the corrections over the Ricci flow due to the higher order terms in the RG flow equation.  We plot these $\delta$  at various order in Fig.\ref{dev expansionsphere} and Fig. \ref{dev expansionhyperbolic}.
It is clear that within the RG flow domain the curvature squared terms
correspond to corrections which can be quantified order by order
in perturbation theory. 

\section{Higher order flow  on warped product manifolds}

\subsection{Geometric flow analysis}

We now consider the behaviour of a five dimensional warped product manifold under the flow Eq.(\ref{RG flow}). The metric on the manifold we consider can be written as,
	\be\label{metric}
		ds^2 = e^{2f\lb( \sigma, \lambda\rb)}\eta_{\mu\nu}dx^\mu dx^\nu + r^2\lb( \sigma, \lambda \rb)d\sigma^2
	\ee
where $\eta_{\mu\nu} = diag[-1,1,1,1]$ is the Minkowski metric.

We will now show that a conformally AdS spacetime, is a solution to the flow upto order $N = 4$ in $\alpha '$. Also we will prove the following propositions --
	\begin{pro}\label{AdS prop}
		The flow Eq.(\ref{RG flow}) with $\beta$ taken upto $O(\alpha '^4)$ has a solution of the form 
		\be\label{AdS soln}
			g(\lambda) = \Omega\lb( \lambda \rb)\lb[ e^{2k\sigma} \eta_{\mu\nu}dx^\mu dx^\nu + d\sigma^2 \rb]
		\ee
		 with $\Omega(\lambda=0) = \Omega_0$ and $k$ is a constant. This is a conformally Anti-deSitter (AdS) spacetime, similar to the bulk 
spacetime in the Randall-Sundrum braneworld model \cite{brane}.
	\end{pro}
	\begin{pro}\label{ODE prop}
		For the solution in Eq.(\ref{AdS soln}) the flow equation upto $O(\alpha '^N)$ where $N \leq 4$ can be reduced to an ODE for $\Omega(\lambda)$ which (after suitable rescaling) is of the form --
		\be\label{RF omega prop}
			\frac{d\Omega}{d\lambda} = \sum_{n=0}^{N-1} a_n\Omega^{-n}
		\ee	
		where $a_n$ are constant coefficients.
	\end{pro}
	\begin{pro}\label{soln 1 3 prop}
		Given $\Omega(\lambda=0)=\Omega_0 \geq 0$, the ODE Eq.(\ref{RF omega prop}) has a unique expanding solution ($\frac{d\Omega}{d\lambda} > 0$), with a finite time singularity in the past at $\lambda = \lambda_s \leq 0$, for each $N = 1,3$ (odd) and also for $N=2,4$ (even) for $\alpha' <0$,
	\end{pro}
	\begin{pro}\label{soln 2 prop}
		For $N=even$ with $\alpha'>0$, Eq.(\ref{RF omega prop}) has $3$ distinct solutions. 
			\begin{itemize}
				\item $\Omega > \alpha'k^2\xi_N$ : An eternal expanding solution without finite time singularities. 
				\item $\Omega < \alpha'k^2\xi_N$ : A contracting solution ending in a singularity at finite $\lambda = \lambda_s \geq 0$.
				\item $\Omega = \alpha'k^2\xi_N$ : A fixed point soliton solution.
			\end{itemize}
		with $\xi_2 = 1$ and $\xi_4 \approx 1.5636 $.
	\end{pro}

	Next we prove these propositions by reducing the flow Eq.(\ref{RG flow}) to an ODE of the form Eq.(\ref{RF omega prop}) and then explicitly solving the ODE.

\subsubsection{\underline{Reduction to ODE}}
	Using the metric of Eq.(\ref{metric}) in the flow Eq.(\ref{RG flow mod}) we get the following dynamical equations for $f$ and $r$ (up to $2nd$ order )
	
\be\label{RF 2 f}
		\dot{f} = \frac{1}{r^2}\lb[ \lb( f'' + f'^2 - \frac{f'r'}{r} \rb) +3f'^2 \rb] - \frac{\alpha '}{r^4}\lb[  \lb( f'' + f'^2 - \frac{f'r'}{r} \rb)^2 +3f'^4 \rb]
	\ee
	\be\label{RF 2 r}
		\dot{r} = \frac{4}{r}\lb[ f'' + f'^2 - \frac{f'r'}{r} \rb] - \frac{4\alpha '}{r^3}\lb[ f'' + f'^2 - \frac{f'r'}{r} \rb]^2
	\ee	

where $\dot{F} = \frac{\partial F}{\partial \lambda}$ and $F' = \frac{\partial F}{\partial \sigma}$, for any function $F$. To solve the above equations we assume that $f$ and $r$ can be written in a variable separable form as --
	\be
		f(\sigma ,\lambda) = f_\sigma (\sigma) + f_\lambda (\lambda) \hspace{0.1in};\hspace{0.1in}  r(\sigma, \lambda) = r_\lambda(\lambda) ~r_\sigma(\sigma)
	\ee
The $r_\sigma$ part can be absorbed into a redefinition of the $\sigma$ coordinate and we essentially have $r = r(\lambda)$. These simplify the equations to give (up to $2nd$ order),

	\be\label{RF 2 f separable}
		\dot{f_\lambda} = \frac{1}{r^2}\lb[ \lb( f_\sigma '' + f_\sigma '^2 \rb) +3f_\sigma '^2 \rb] - \frac{\alpha '}{r^4}\lb[  \lb( f_\sigma '' + f_\sigma '^2  \rb)^2 +3f_\sigma '^4 \rb]
	\ee
	\be\label{RF 2 r separable}
		\dot{r} = \frac{4}{r}\lb[ f_\sigma '' + f_\sigma '^2 \rb] - \frac{4\alpha '}{r^3}\lb[ f_\sigma '' + f_\sigma '^2  \rb]^2
	\ee	

Since the equations are now in a variable separable form, for consistency at the leading order we must have $f_\sigma '' + f_\sigma '^2 = const_1$ and $\lb( f_\sigma '' + f_\sigma '^2 \rb) +3f_\sigma '^2 = const_2$; i.e. $f_\sigma ' = k = const.$ and $f_\sigma = k\sigma$. Note that this condition comes only from the leading order terms. The equations in $\lambda$ become $	\frac{r^2}{e^{2f_\lambda}}\dot{\lb(e^{2f_\lambda}\rb)} = \dot{\lb(r^2\rb)}$, which gives $e^{2f_\lambda} = r^2 = \Omega(\lambda) \geq 0$.\\
Thus we have a solution with the metric Eq.(\ref{AdS soln}) which is conformal to the Anti-de Sitter metric, as stated in Prop.\ref{AdS prop}. The Ricci and Kretschmann scalars, $R= g^{ij}R_{ij}$ and $K= R_{ijkl}R^{ijkl}$ respectively, for such a metric are given by --

	\be\label{R scalar}	
		R  = - \frac{4}{r^2}\left[ 2f''+5f'^2\right] = -\frac{20k^2}{\Omega}
	\ee
	\be\label{K scalar}
		K = \frac{8}{r^4}\lb[ 2\lb( f'' + f'^2 \rb)^2 + 3f'^4 \rb] = \frac{40k^4}{\Omega^2}
	\ee
Thus the curvature diverges and the manifolds become singular at $\Omega = 0$.

Using the separable form for the functions in Eq.(\ref{beta 3}), we get 
(at third order),
	\be\label{beta 3 separable}
		\begin{split}
			\beta^{(3)}_{\mu\mu} & = \frac{e^{2f}}{4r^6}\lb(-32f_\sigma '^6 - 16f_\sigma '^4f_\sigma '' - 12f_\sigma '^2f_\sigma ''^2 - 14f_\sigma ''^3 + 4f_\sigma '^3f_\sigma ^{(3)} + 8f_\sigma 'f_\sigma ''f_\sigma ^{(3)} + {f_\sigma ^{(3)}}^2 \rb) \\
			\beta^{(3)}_{\sigma\sigma} & = \frac{1}{r^4}\lb(-8f_\sigma '^6 - 12f_\sigma '^4f_\sigma '' + 3f_\sigma '^2f_\sigma ''^2 - 3f_\sigma ''^3 + 5f_\sigma '^3f_\sigma ^{(3)} + 6f_\sigma 'f_\sigma ''f_\sigma ^{(3)} + {f_\sigma ^{(3)}}^2 + f_\sigma '^2f_\sigma ^{(4)} + f_\sigma ''f_\sigma ^{(4)}\rb) \\	
		\end{split}
	\ee
where $F^{(n)} = \frac{d^n F}{d\sigma^n}$. But, from the separability at leading order, we already have $f' = k$ and thus both the terms reduce to $e^{-2f}r^6\beta^{(3)}_{\mu\mu} = r^4\beta^{(3)}_{\sigma\sigma} = -8k^6$. Similarly all higher derivative terms vanish at order $4$ leaving $e^{-2f}r^8\beta^{(4)}_{\mu\mu} = r^6\beta^{(4)}_{\sigma\sigma} = 2 \lb( 3+5\zeta(3)\rb)k^8$. 

This leads to the following ODE for $\Omega$ --
	
	\be\label{RF Omega}
		\frac{1}{8k^2}\frac{d\Omega}{d\lambda} =  1 - \frac{\alpha 'k^2}{\Omega} + 2 \lb(\frac{\alpha 'k^2}{\Omega}\rb)^2 - \frac{3+5\zeta(3)}{2} \lb(\frac{\alpha 'k^2}{\Omega}\rb)^3
	\ee

We can readily see that $k=0$ is a fixed point of the flow. This corresponds to a 5 dimensional Minkowski space which is thus, a \emph{soliton} of the flow as is expected. For $k \neq 0$ we rescale the variables using --

	\be\label{bar defn}
		\bar\Omega = \frac{\Omega}{\vert\alpha'\vert k^2} \hspace{20pt};\hspace{20pt} \bar\lambda = \frac{8\lambda}{\vert\alpha'\vert}
	\ee
with $\bar\Omega \geq 0$ and the Eq.(\ref{RF Omega}) becomes --
	
	\be\label{RF Omega bar}
		\frac{d\bar\Omega}{d\bar\lambda} =  1 \mp \bar\Omega^{-1} + 2 \bar\Omega^{-2} \mp \frac{3+5\zeta(3)}{2} \bar\Omega^{-3}
	\ee

Here, the upper signs are for $\alpha' >0$ and the lower ones for $\alpha ' < 0$, a convention we shall adopt in all that follows. For either case this equation is of the form of Eq.(\ref{RF omega prop}) in Prop.(\ref{ODE prop}). Note that now a term of the form $\bar \Omega ^{n-1}$ corresponds to $O(\alpha'^n)$ in the original flow Eq.(\ref{RG flow}).

\subsubsection{\underline{Solitons, Solutions and Singularities}}
We can solve Eq.(\ref{RF Omega bar}) for all orders and obtain explicit solutions.

At $N =1$ order, the flow is essentially an \emph{un-normalized} Ricci flow independent of $\alpha '$ with the solution --
	\be\label{Omega soln 1}
		\bar\lambda + C_1 = \bar\Omega
	\ee

For $N =2$ there exists a fixed point soliton for $\bar{\Omega} = \xi_2 = 1$ and when $\bar{\Omega} \neq 1$ we get -- 
	\be\label{Omega soln 2}
		\bar{\lambda} + C_2 = \bar{\Omega} \pm \ln \lb\vert \bar{\Omega} \mp 1 \rb\vert
	\ee

Order $N=3$ has no fixed points and the solution is --

	\be\label{Omega soln 3}
		\bar\lambda + C_3 = \bar\Omega \pm \frac{1}{2}\ln \lb\vert \bar\Omega^2 \mp \bar\Omega + 2 \rb\vert - \frac{3}{\sqrt{7}}\tan^{-1}\lb( \frac{2\bar\Omega \mp 1}{\sqrt{7}} \rb)
	\ee\\

The ODE at $N=4$ has one fixed point at $\bar\Omega = \xi_4 \approx 1.5636$. To obtain the solutions for $\bar\Omega \neq \xi_4$ we use the following factorization and partial fraction splittings --

	\begin{subequations}\label{math trick}
		\begin{align}
 			\bar\Omega^3 - \bar\Omega^2 + 2\bar\Omega - \frac{3+5\zeta(3)}{2} & = \lb( \bar\Omega - \xi_4 \rb)\lb( \bar\Omega^2 + \beta \bar\Omega +\gamma \rb) \label{factor}\\
			\frac{\bar\Omega^2 - 2\bar\Omega + \frac{3+5\zeta(3)}{2}}{\lb( \bar\Omega - \xi_4 \rb)\lb( \bar\Omega^2 + \beta \bar\Omega +\gamma \rb)} & = \frac{a}{\bar\Omega - \xi_4} - \frac{b \bar\Omega + c}{\bar\Omega^2 + \beta \bar\Omega +\gamma} \label{partial frac}
		\end{align}
	\end{subequations}

where the various parameters can be found approximately as --	
	\begin{subequations}\label{math trick param}
		\begin{align}
 			\xi_4 \approx 1.5636 \hspace{10pt} & \beta \approx 0.5636 \hspace{10pt} \gamma \approx 2.8812 \label{factor param}\\
			a \approx 0.6158 \hspace{10pt} & b \approx -0.3841 \hspace{10pt} c \approx 1.7464 \label{partial frac param}
		\end{align}
	\end{subequations}

In terms of these parameters the solution for $N=4$ can be written as --

	\be\label{Omega soln 4}
		\bar\lambda + C_4 = \bar\Omega \pm a \ln \lb\vert \bar\Omega \mp \xi_4 \rb\vert \mp \frac{b}{2}\ln \lb\vert \bar\Omega^2 \pm \beta \bar\Omega + \gamma \rb\vert - \frac{2 c - \beta b}{\sqrt{4\gamma - \beta^2}}\tan^{-1}\lb( \frac{2\bar\Omega \pm \beta}{\sqrt{4\gamma - \beta^2}} \rb)
	\ee

\begin{figure} 
\centering
\subfigure[$\bar\Omega_0= 0.5$]{\includegraphics[width=0.4\textwidth]{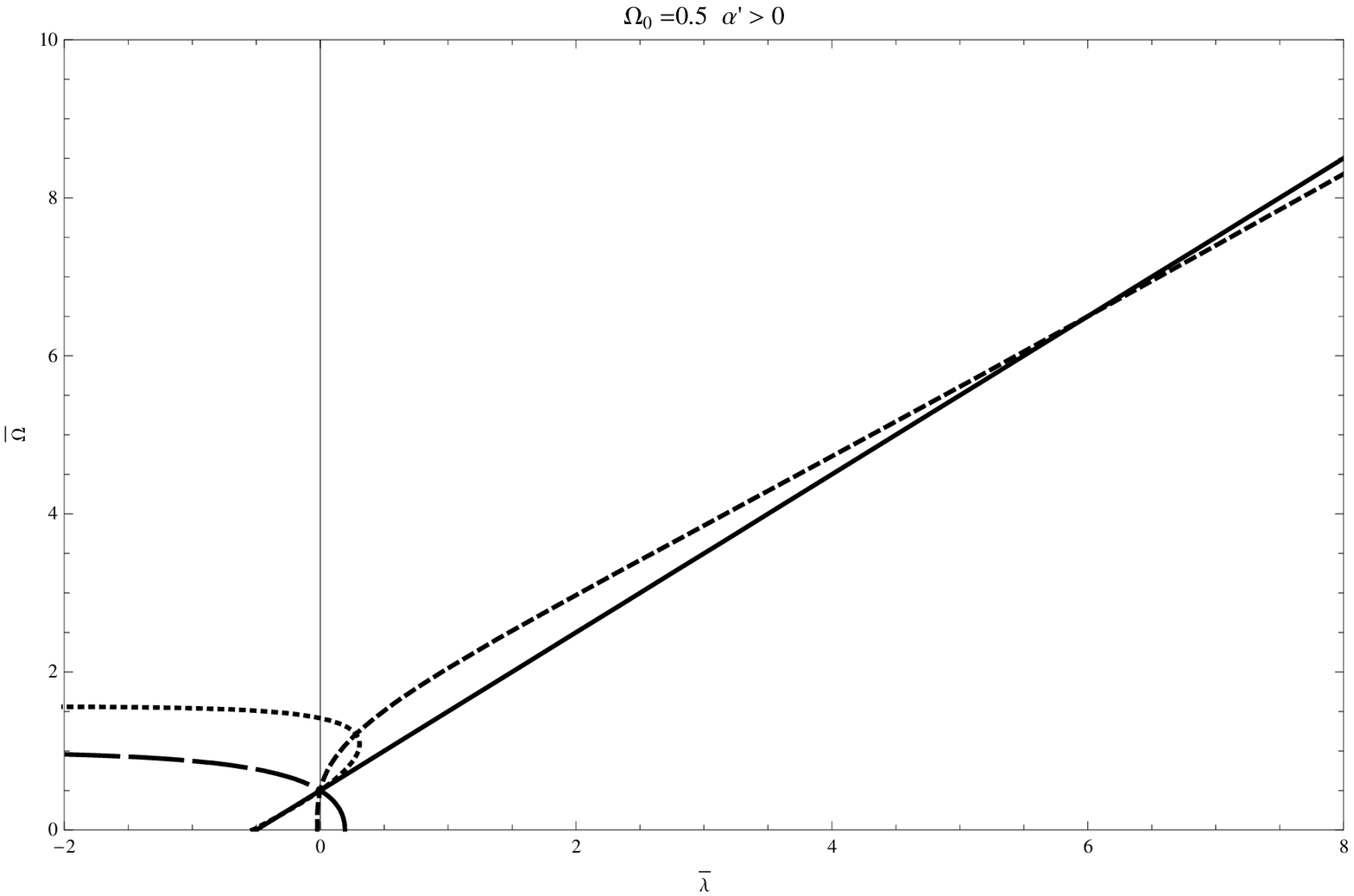} \includegraphics[width=0.4\textwidth]{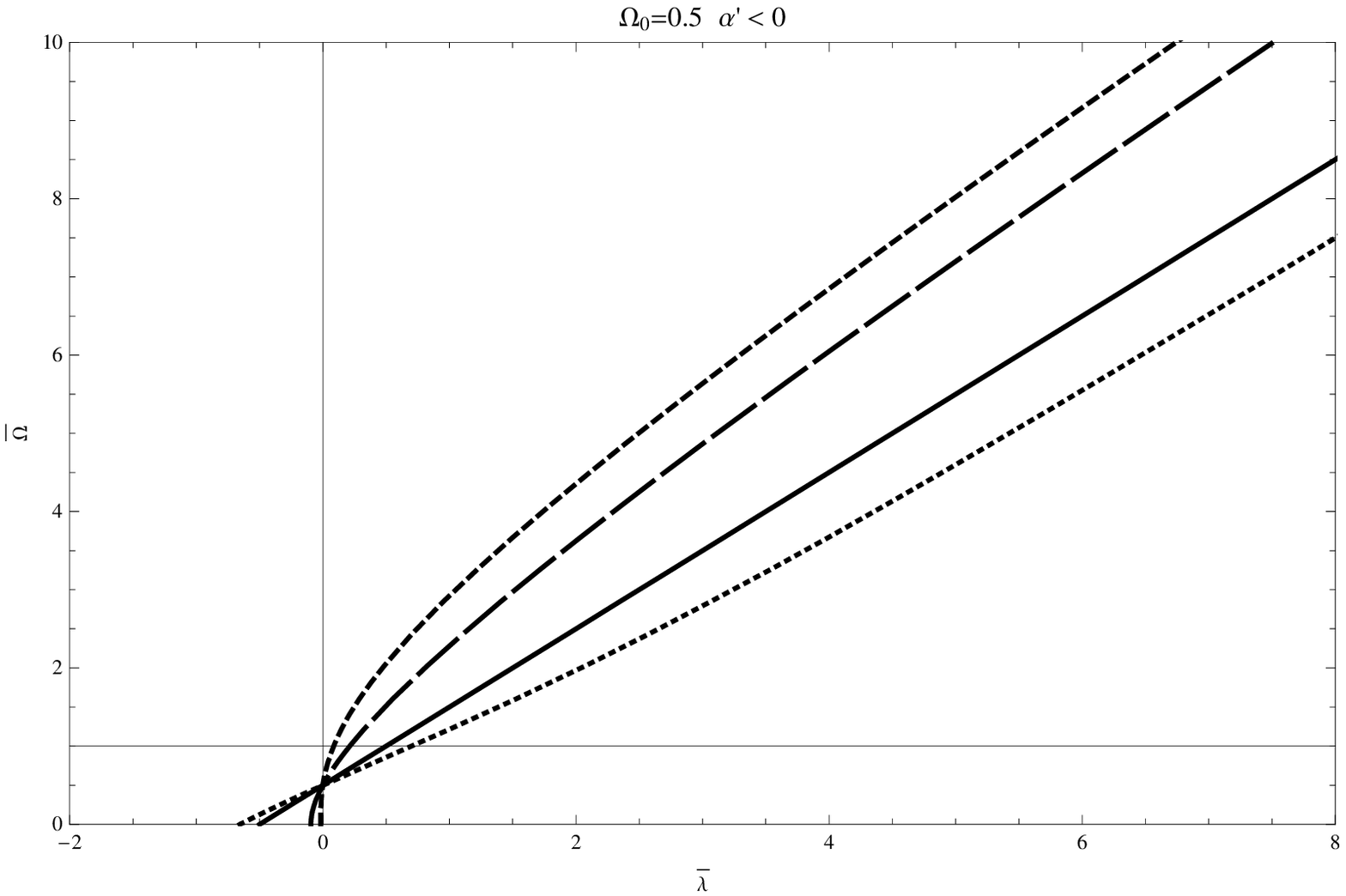}}
\subfigure[$\bar\Omega_0= 1$]{\includegraphics[width=0.4\textwidth]{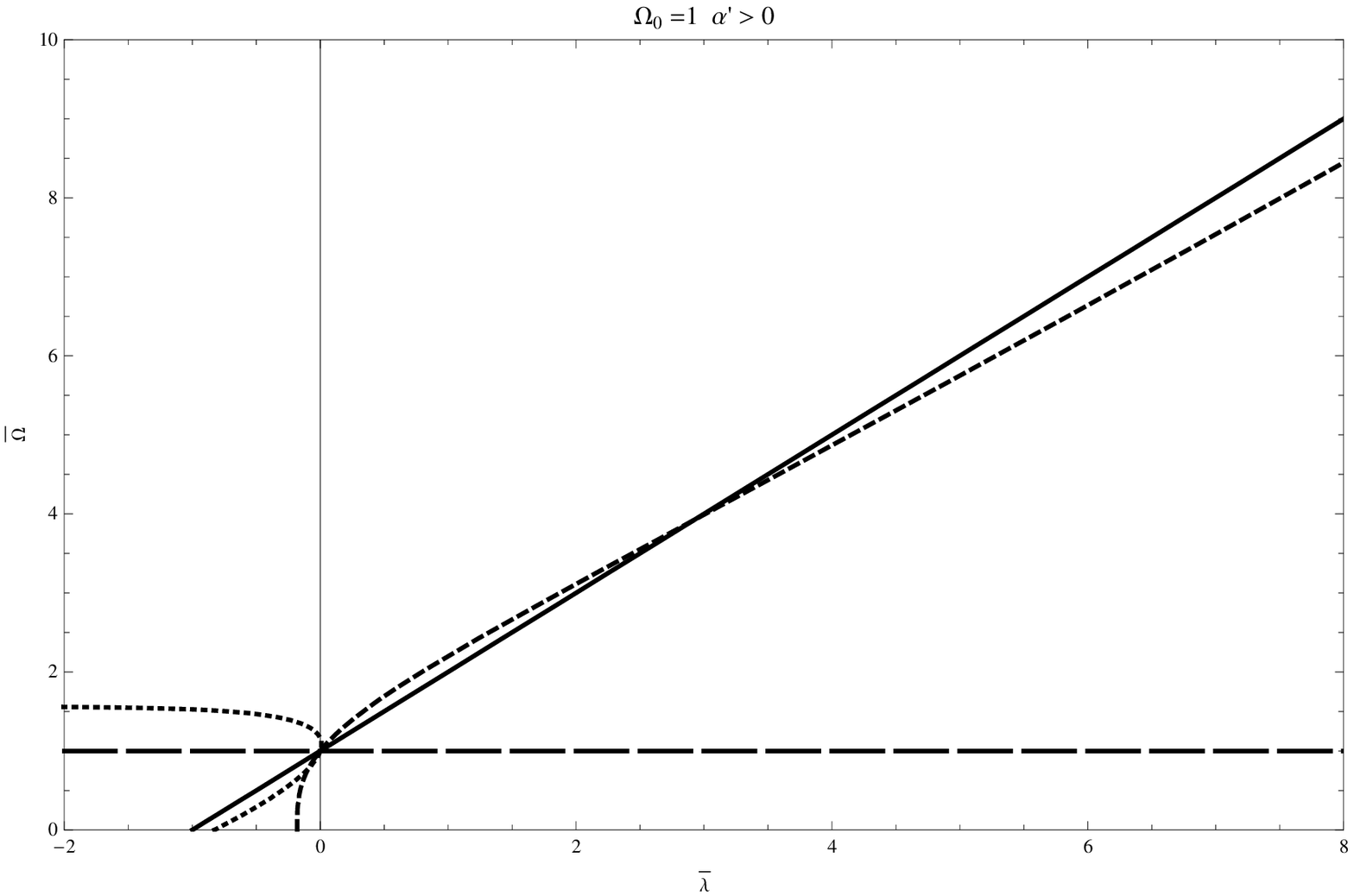} \includegraphics[width=0.4\textwidth]{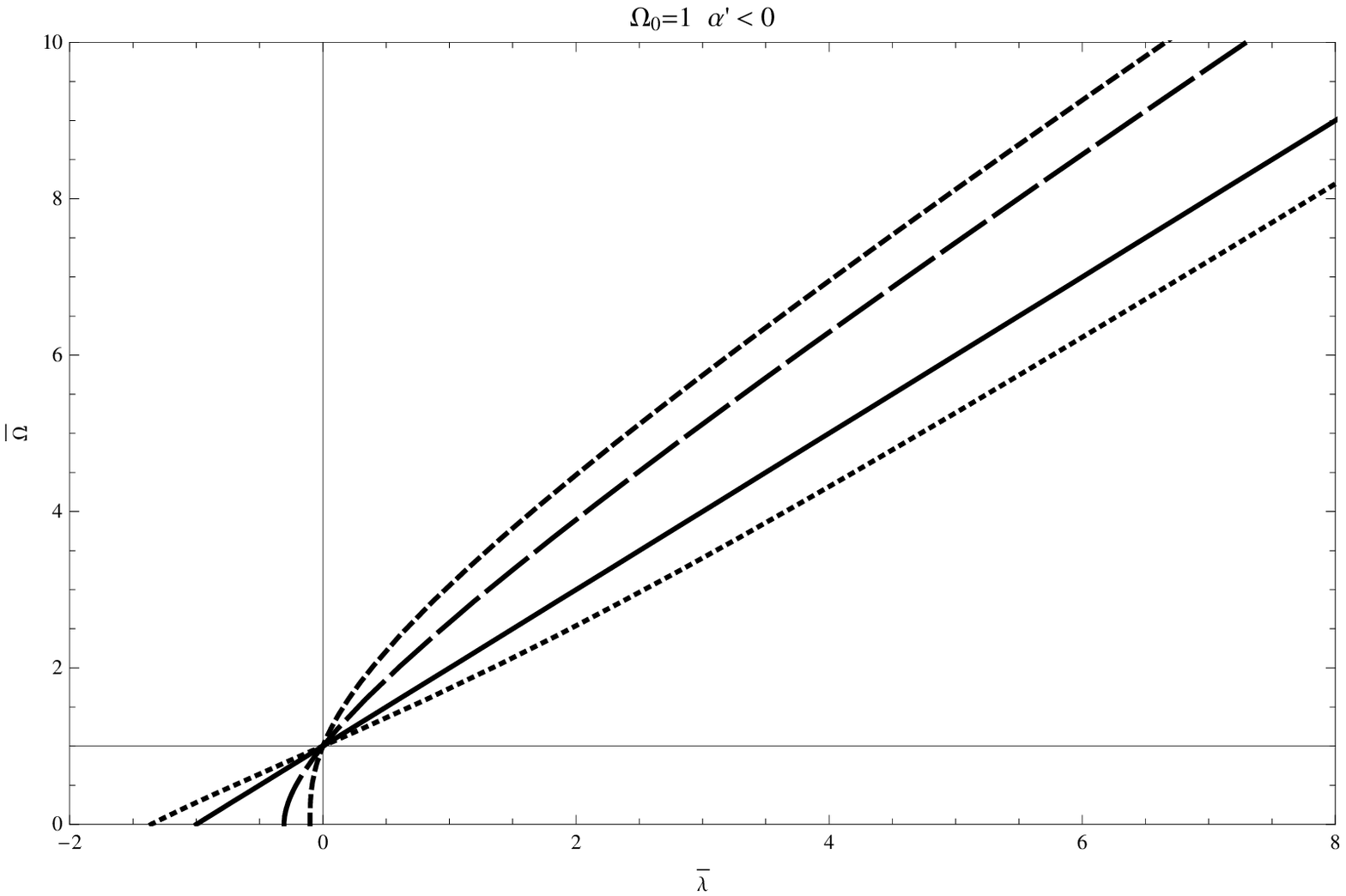}}
\subfigure[$\bar\Omega_0=  1.5636 $]{\includegraphics[width=0.4\textwidth]{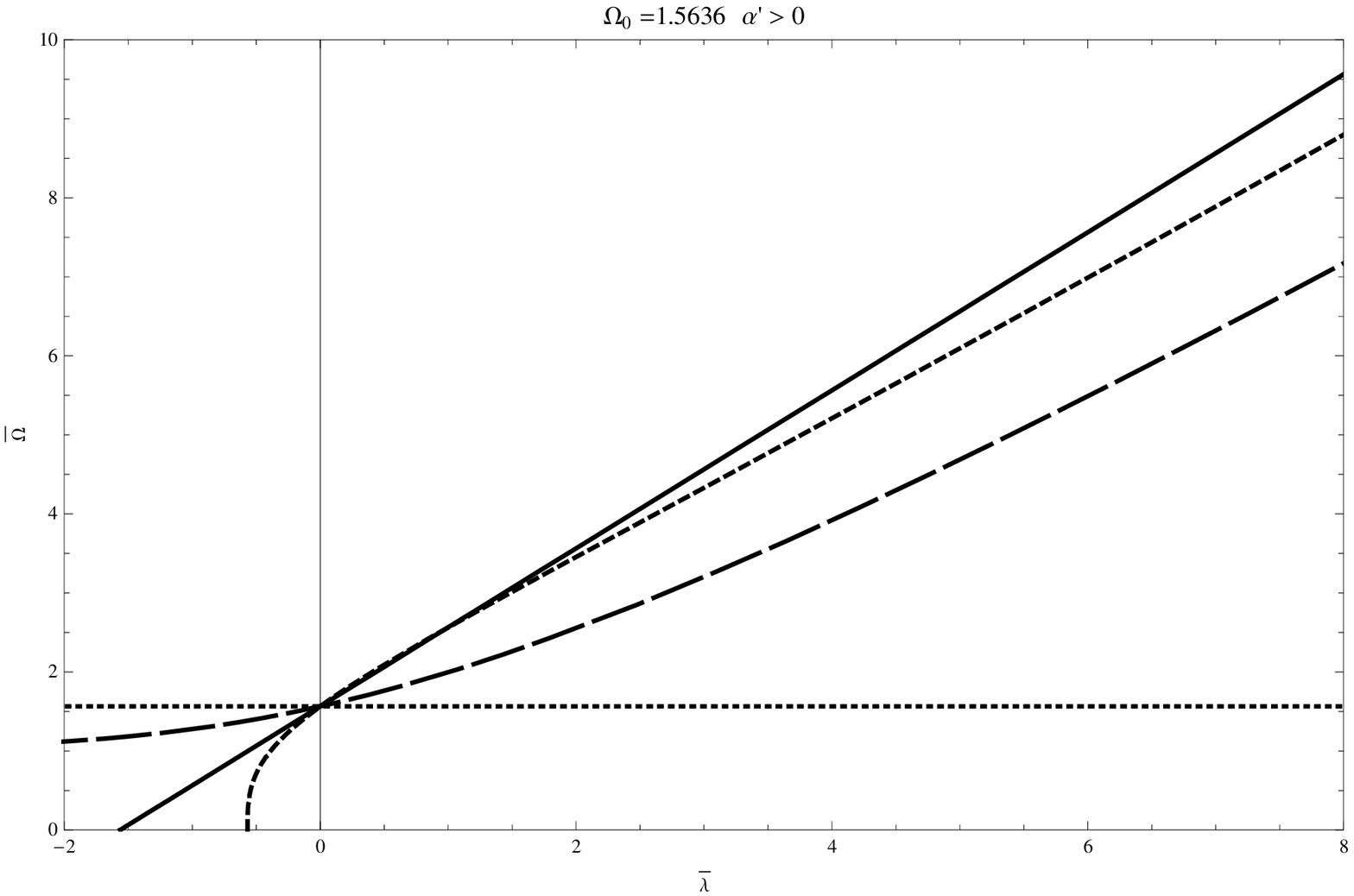} \includegraphics[width=0.4\textwidth]{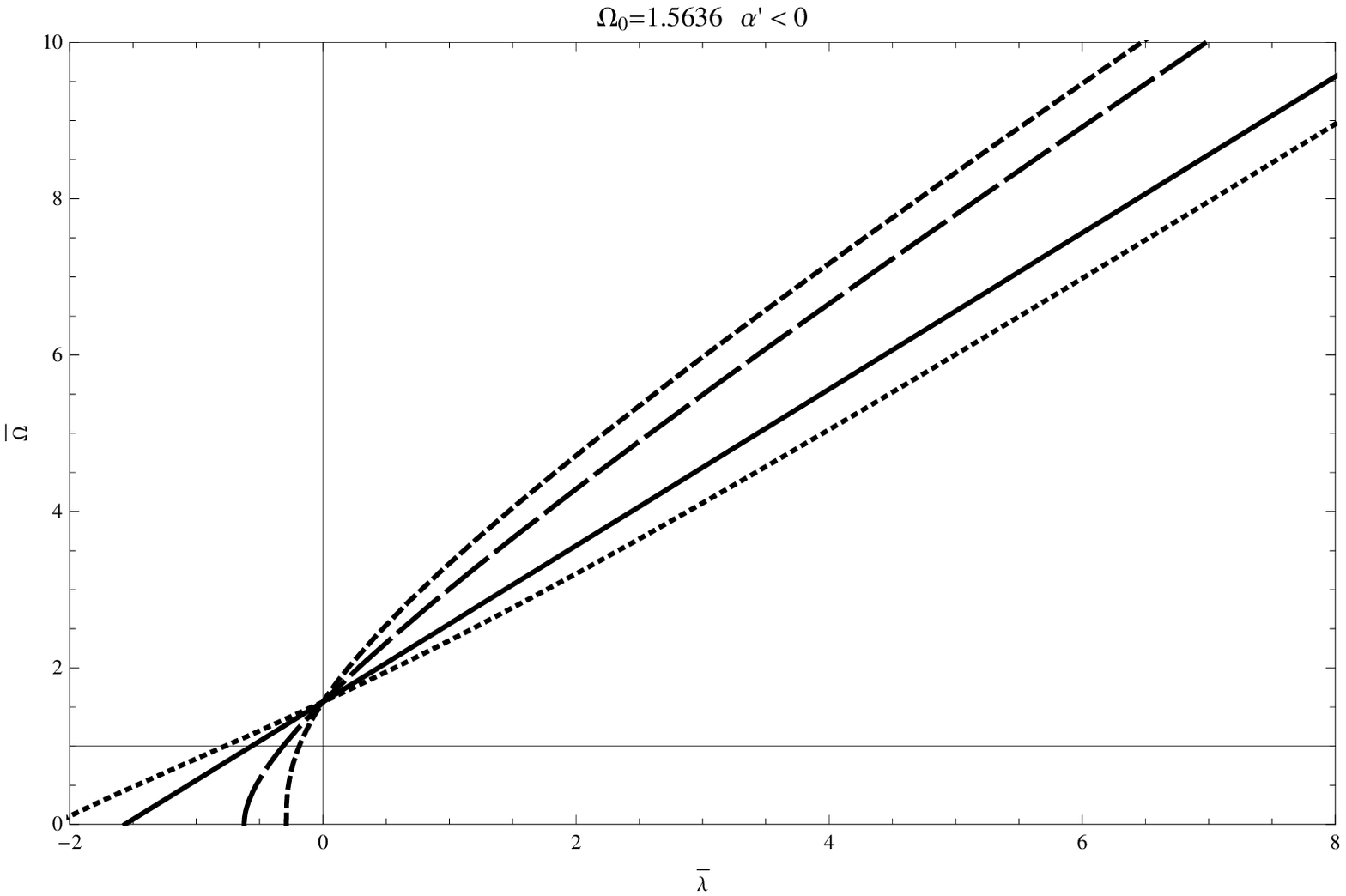}}
\subfigure[$\bar\Omega_0= 4$]{\includegraphics[width=0.4\textwidth]{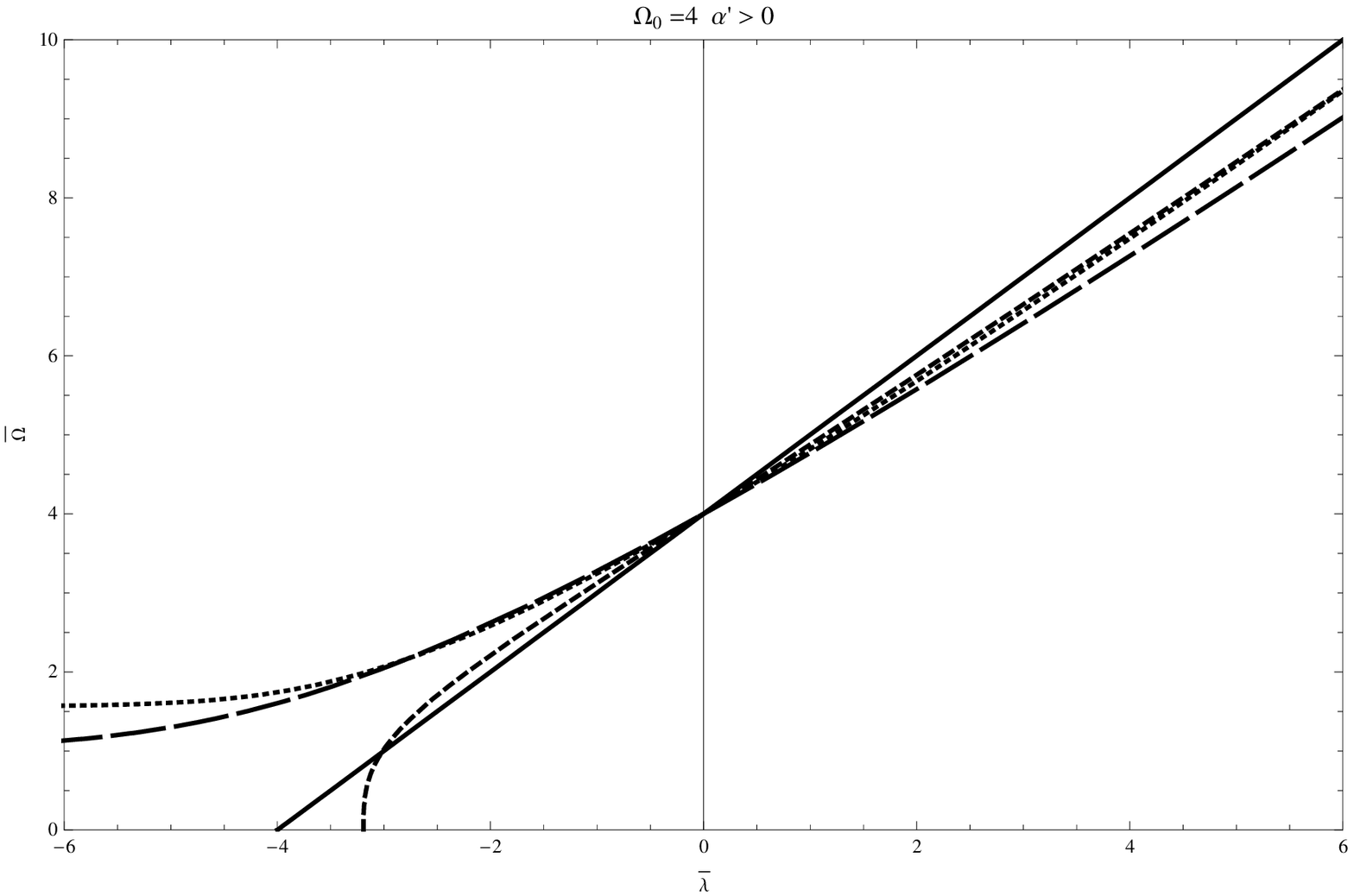} \includegraphics[width=0.4\textwidth]{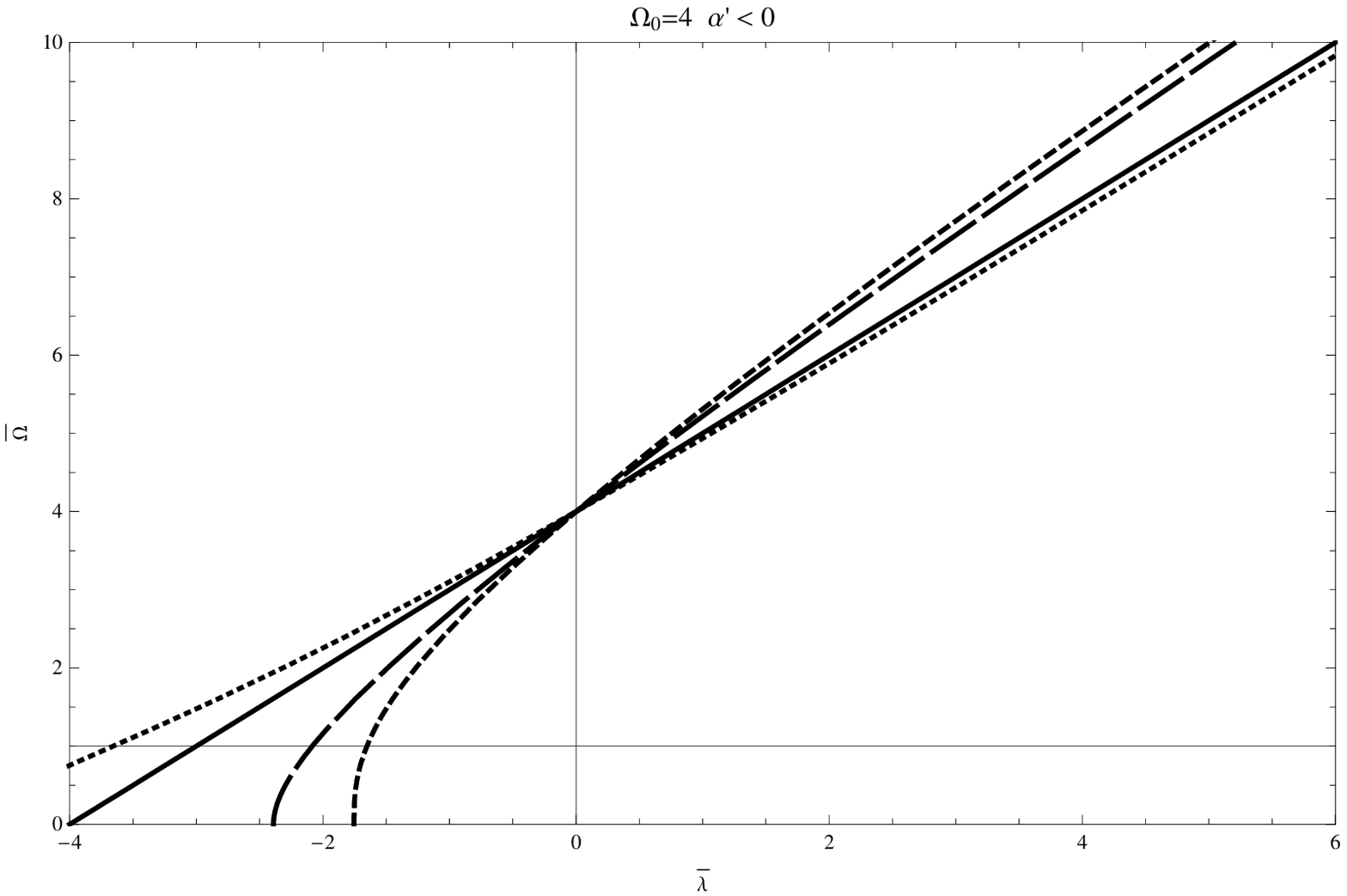}}
\caption{$\bar\Omega$ vs $\bar\lambda$ for various inital values of $\bar\Omega_0$ at $\bar\lambda = 0$. The solutions at different orders are $1^{st}$ (continuous line), $2^{nd}$ (long dashed line), $3^{rd}$ (short dashed line) and $4^{th}$ (dotted line). The figures on the left are for $\alpha'>0$ and on the right are $\alpha' < 0$}
\label{omega lambda}
\end{figure}

To compare these solutions, we plot all of them with the same initial condition $\bar\Omega (\bar\lambda = 0) = \bar\Omega_0$ in Fig.{\ref{omega lambda}}. 

It can be seen that the $N=odd$ order solutions and the $N=even$ solution for $\alpha '< 0$ begin at a finite time singularity and continue to expand indefinitely. Whereas for $N=even$ order and $\alpha' >0$ we have 3 solutions - a soliton ($\bar\Omega = \xi_N$), an eternal solution ($\bar\Omega > \xi_N$) and a solution with a finite time singularity ($\bar\Omega < \xi_N$).

\begin{figure}
\centering
\subfigure[$\alpha' > 0$]{\includegraphics[width=0.4\textwidth]{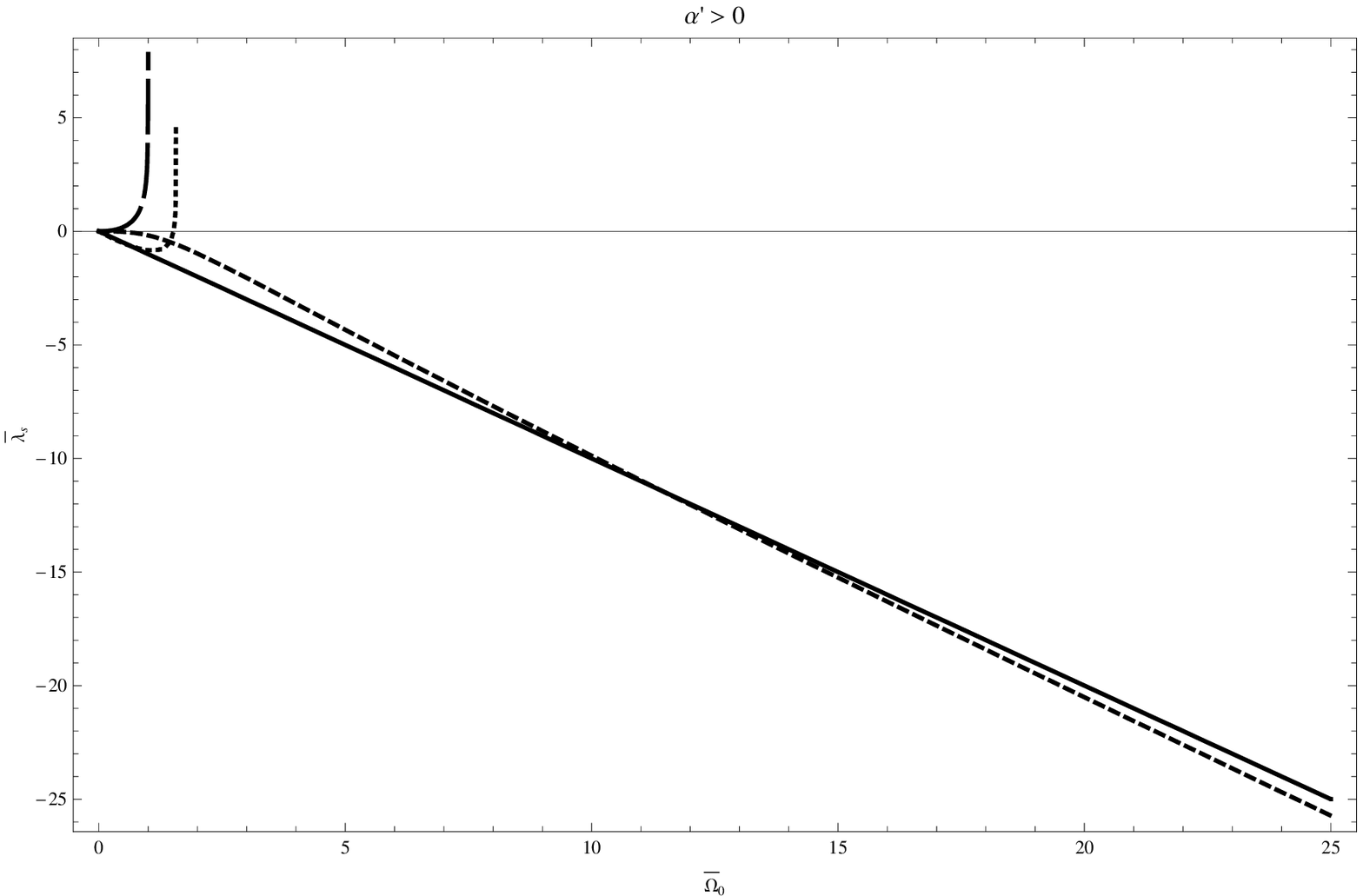}\label{ls>0}}
\subfigure[$\alpha' < 0$]{\includegraphics[width=0.4\textwidth]{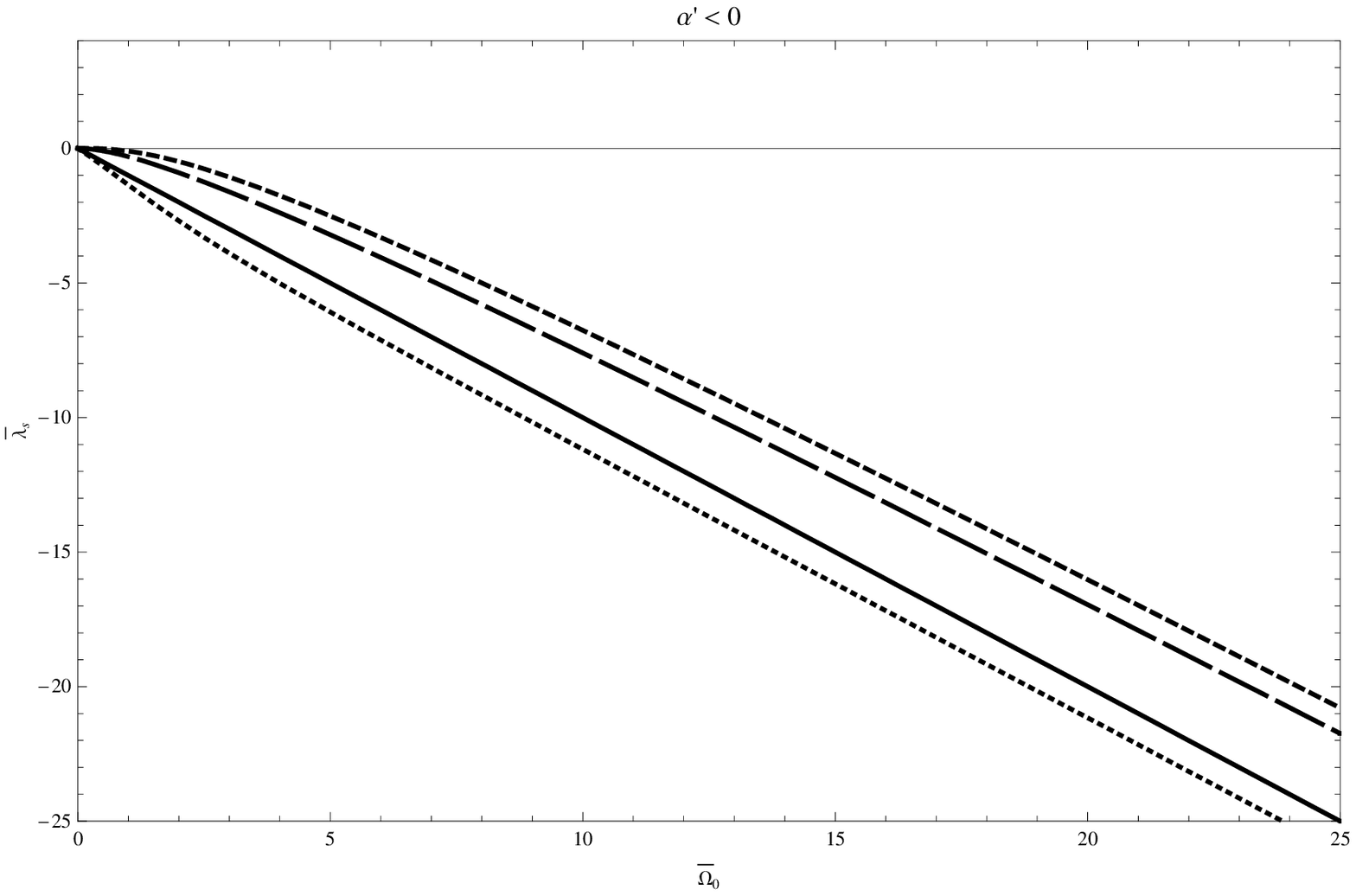}\label{ls<0}}
\caption{$\bar\lambda_s$ vs $\bar\Omega_0$. The plots for different orders are $1^{st}$ (continuous line), $2^{nd}$ (long dashed line), $3^{rd}$ (short dashed line) and $4^{th}$ (dotted line).}
\label{sing time}
\end{figure}

The singularity time $\lambda=\lambda_s$ such that $\Omega(\lambda_s) = 0$ depends on $\Omega_0$. This behaviour can be seen in the Fig.\ref{sing time}.

We see from Fig.\ref{sing time} that $\bar\lambda_s \leq 0$ except for the case of order $N=even$ and $\alpha'>0$ in which case $\bar\lambda_s$ starts from zero and then increases upto infinity as $\bar\Omega_0 \to \xi_N$. Furthermore for $\alpha' <0$ we always have $\bar\lambda_s^{(4)} <\bar\lambda_s^{(1)} < \bar\lambda_s^{(2)} < \bar\lambda_s^{(3)}$ for the respective orders of RG flow as seen in Fig.\ref{ls<0}.

\subsection{Perturbative RG flow domain}
Considered as an RG flow equation, Eq.(\ref{RF Omega bar}) is valid only in the region where $1/\bar\Omega \ll 1$. Note that this regime does not contain any solitons or singularities. To study the flow in that limit, we expand the solutions obtained at various orders Eqs.(\ref{Omega soln 1})-(\ref{Omega soln 3}) and (\ref{Omega soln 4}) in the powers of $1/\bar\Omega$ as follows --

	\begin{subequations}
		\begin{align}
			\lambda + C_1 & =  \bar\Omega \\
			\lambda + C_2 & = \bar\Omega \pm \ln\bar\Omega - \lb( \frac{1}{\bar\Omega} \rb) \mp \frac{1}{2} \lb( \frac{1}{\bar\Omega} \rb)^2 - \frac{1}{3} \lb( \frac{1}{\bar\Omega} \rb)^3 + \ldots \\
			\lambda + \tilde C_3 & =  \bar\Omega \pm \ln\bar\Omega + \lb( \frac{1}{\bar\Omega} \rb) \pm \frac{3}{2} \lb( \frac{1}{\bar\Omega} \rb)^2 + \frac{1}{3} \lb( \frac{1}{\bar\Omega} \rb)^3 + \ldots\\
			\lambda + \tilde C_4 & =  \bar\Omega \pm \ln\bar\Omega + \lb( \frac{1}{\bar\Omega} \rb) \mp 0.7526 \lb( \frac{1}{\bar\Omega} \rb)^2 - 2.6699 \lb( \frac{1}{\bar\Omega} \rb)^3 + \ldots	
		\end{align}
	\end{subequations}
where constant terms have been absorbed as $\tilde C_3 = C_3 + \frac{3\pi}{2\sqrt{7}}$ and $\tilde C_4 = C_4 + 1.74045$.

\begin{figure}
\centering
\subfigure[$\alpha' > 0$]{\includegraphics[width=0.4\textwidth]{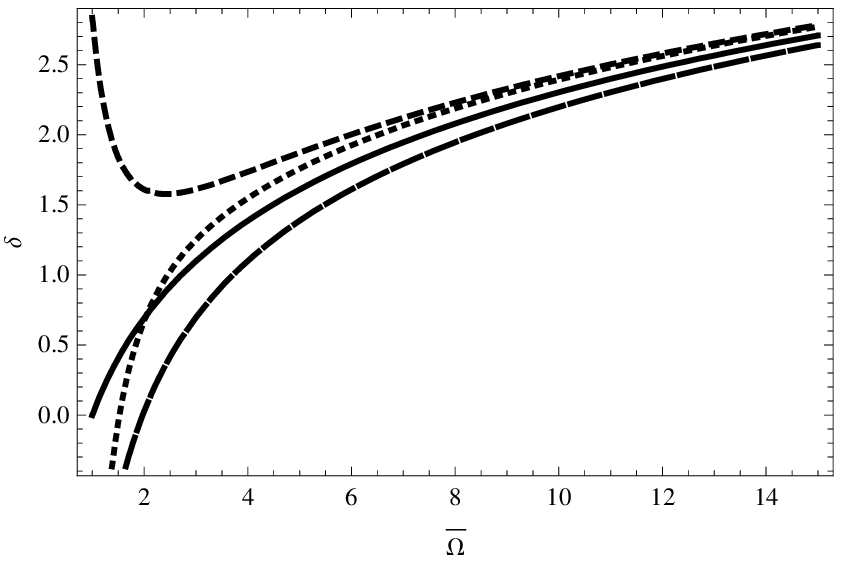}\label{dev>0}}
\subfigure[$\alpha' < 0$]{\includegraphics[width=0.4\textwidth]{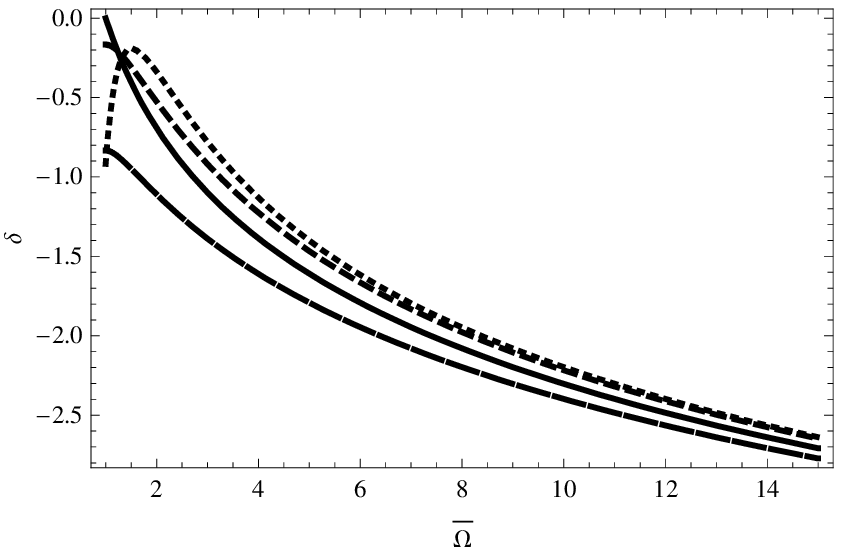}\label{dev<0}}
\caption{$\bar\lambda_s$ vs $\bar\Omega_0$. The plots for $\pm\ln\bar\Omega$ (continuous line) and $\delta$ at different orders, $2^{nd}$ (long dashed line), $3^{rd}$ (short dashed line) and $4^{th}$ (dotted line).}
\label{dev expansion}
\end{figure}

We can write all pof these as $\lambda + C = \bar\Omega + \delta$ where $\delta$ represents the corrections over the Ricci flow due to the higher order terms in the RG flow equation.  We plot these $\delta$ over the Ricci flow obtained at various order in Fig.\ref{dev expansion}. We see that the leading correction over the Ricci flow is $\sim \ln\bar\Omega$, and other higher order corrections then vanish in the limit of large $\bar\Omega$.

We can see that the $2^{nd}$ order solution is correct upto $\ln\bar\Omega$ term and the $3^{rd}$ order upto $1/\bar\Omega$. We suspect, similarly, that the $4^{th}$ order solution will be correct till $1/\bar\Omega^2$, but this can be verified only if even higher order solutions are available.

\section{Summary and Conclusion}

In this article we have investigated the higher order flows on various 
manifolds, from uniform homogeneous spaces to warped manifolds. Let us summarize the work done and mention some future possibilities briefly.
  
We began with higher order geometric flows for the toy examples of the two 
sphere and hyperbolic space in two dimensions. It turns out that the results for the
1st order (Ricci) and 3rd order flows are similar in nature while the
2nd and 4th order flows have similarities. There is a marked difference
for flows with $\alpha'>0$ and $\alpha'<0$, which has been pointed out.
The existence of fixed points (solitons) have also been mentioned. 
We emphasize that these toy examples, though simplistic, do provide us
with pointers towards distinguishing between the specific effects
that occur because of the inclusion of higher order terms. As expected,
the higher order terms do not {\em resolve} the lower order singularities
though there are definite changes in terms of the singularity `time',
initial condition dependencies and the appearance of solitons.  
Further, we discuss the perturbative RG flow domain. Here, we are able to
quantify the corrections due to the presence of the higher order terms.

Subsequently, we turned towards higher order flows in the context of
warped product manifolds. This analysis is largely inspired by
the physics of warped extra dimensions and brane--bulk models. 
As claimed in Prop.\ref{AdS prop} we show that the conformally AdS spacetime 
in Eq.(\ref{AdS soln}) is a solution of the RG flow upto order $N=4$ in 
$\alpha'$. While reducing the RG flow PDEs to ODEs, assuming separability of 
the functions in the line element, we noted that the conformal AdS nature 
arises by using just the leading order terms. The higher order equations are 
then consistent, giving the conformal factor $e^{2f_\lambda} = r^2 = \Omega$. 
The behavior of $\Omega$ is then determined by the ODE and the order of the 
expansion of the beta functions. 

It is then natural to conjecture that the conformally AdS metric 
Eq.(\ref{AdS soln}) will solve the flow upto any order $N$. Note that 
using only the leading order we get $f'_\sigma = k$. Also since curvature 
is a function of derivatives of the metric, the $\beta
$-function at any order $N$ will look like $\beta^{(N)}_{\mu\mu} \sim \frac{e^{2f}A^{(N)}(k)}{r^{2N}}$ and $\beta^{(N)}_{\sigma\sigma} \sim \frac{B^{(N)}(k)}{r^{2(N-1)}}$, where $A$ and $B$ are functions of $k$, just like in Eq.({\ref{beta 3 separable}}). If $A^{(N)}(k)= B^{(N)}(k)$ then we can write $e^{2f} = r^2=\Omega$ and thus get the conformally AdS solution. Then the ODE for $\Omega$ 
will also be of the form of Eq.(\ref{RF omega prop}) for any order $N$. 
Thus we conjecture that both Prop.\ref{AdS prop} and Prop.\ref{ODE prop} 
hold for any order $N$. Since explicit forms of the $\beta_{ij}$ are not 
known for $N>4$, we cannot explicitly check these conjectures presently. 
It will be interesting to look for a proof using general principles.

Apart from the Minkowski spacetime, we have found another soliton solution to the flow at even orders $N=2,4$. For this solution $\Omega = \alpha'k^2\xi_N$ and from Eq.(\ref{R scalar}) we see that $R = -20/\alpha'\xi_N$. Thus the spacetime has curvature inversely proportional to $\alpha'$. In the context of RG flows,
the existence of a soliton is a non-perturbative effect. 
Smaller values of $\alpha'$ correspond to larger curvatures and hence the 
perturbative expansion of the $\beta$ functions in Eq.(\ref{beta expansion}) 
fails. It is for this reason that the solutions obtained at various orders 
show markedly different behaviour close to the fixed point value of 
$\bar\Omega = \xi_N$. In regions of low curvature, we have evaluated the
higher order corrections to the RG flows--these corrections are small
as long as we are in the perturbative RG flow domain. However, they are
useful in quantifying the RG flow behaviour at successive higher orders. 
Moreover, the singularities and solitons which appear in the geometric
flow analysis can serve as pointers to the precise domain in which the
$\sigma$-model RG flow equations are valid.  

In a way, we have shown that the bulk spacetime used in the warped braneworld models naturally arise as solutions of the flow equations (atleast upto
fourth order). Though this result assumes separability of the metric functions we are able to demonstrate the uniqueness of AdS spacetime as a solution of the 
flow equations. The task of looking at non--separable situations, which at the lowest order was discussed in our previous article \cite{dpk} remains and will be taken up in future. 

Further, the evolution of geometric quantities such as the Riemann, Ricci scalar in the context of higher order flows will be an exercise worth
pursuing--the primary motivation being to distinguish between the lowest order(Ricci) and these higher order geometric flows.

%\end{thebibliography}


\begin{references}
\bibitem{friedan} D. ~Friedan, Phys. Rev. Letts. {\bf 45} 1057 (1980), 
D.~ Friedan, Annals of Physics {\bf 163}, 318 (1985).
\bibitem{sen} A. Sen, Phys. Rev. Lett. {\bf 55}, 1846 (1985); C. G. Callan, E. T. Martinec, M. T. Perry and D. Friedan, Nucl. Phys. {\bf B262}, 593 (1985).
\bibitem{hamilton} R.S.~ Hamilton, J. Diff. Geom. {\bf 17}, 255 (1982).
\bibitem{perelman} G. Perelman,The entropy formula for the Ricci flow and its geometric applications,  Preprint math.DG/0211159. 
\bibitem{Chow} B.~Chow and D.~Knopf, \emph{The Ricci flow: an
introduction},  Mathematical Surveys and Monographs Vol. 110, AMS, 
Providence, 2004.
\bibitem{polchinski} J. Polchinski, {\em String Theory, volume I}, Cambridge University Press, Cambridge, UK, 1998.  
\bibitem{oliy} T. Oliynyk, V. Suneeta and E. Woolgar, Phys. Lett. {\bf B610}, 115 (2005); T. Oliynyk, V. Suneeta and E. Woolgar, Nucl. Phys. {\bf B739}, 441 (1006); J. D. Streets, J. Geom. Phys. {\bf 59}, 8 (2009).
\bibitem{carfora} M. Carfora, Renormalization group and the Ricci flow,
arXiv:1001.3595.
\bibitem{bakas} I. Bakas, Renormalization group equations and geometric flows,
arXiv:hep-th/0702034.
\bibitem{oli} T.~Oliynyk, V.~Suneeta and E. Woolgar, Phys.Rev.{\bf D76},045001
(2007).
\bibitem{oli1} T.~Oliynyk, Class. Quantum Grav. {\bf 26}  105020, (2009);  C. Guenther, and T. A. Oliynyk, Lett. Math. Phys. {\bf 84}, 149 (2008). 
\bibitem{streets} J. D. Streets, J. Geom. Anal.{\bf 18},249 (2008).
\bibitem{tseytlin} A. A. Tseytlin, Phys. Rev.{\bf D75}, 064024 (2007). 
\bibitem{brane} L. Randall and R. Sundrum, Phys. Rev. Lett.{\bf 83},4690 (1999);  Phys. Rev. Lett. {\bf 83}, 3370(1999).
\bibitem{jack} I. Jack,  D. R. T. Jones, and N. Mohammedi,  Nuc. Phys. B{\bf 322}  (1989), 431-470. 
\bibitem{dpk} S. Das, K. Prabhu and S. Kar, Ricci flow of unwarped and warped product manifolds, arXiv:0908.1295 (to appear in IJGMMP).


\end{references}
\end{document}